\documentstyle[psfig]{mn}

\footnotesize
\newdimen\digitwidth    %define ! a one digit width for tables
\setbox0=\hbox{\rm0}
\digitwidth=\wd0
\catcode`!=\active
\def!{\kern\digitwidth}
\normalsize

\title[Young Pulsars \& Discovery and Timing of 200 Pulsars] {The Parkes multibeam
pulsar survey: III. Young pulsars \& the 
discovery and timing of 200 pulsars}
\author[M. Kramer et al.]
{
M. Kramer$^1$,\thanks{Email: mkramer@jb.man.ac.uk}
J.~F. Bell$^2$,
R.~N. Manchester$^2$,
A.~G. Lyne$^1$,
F. Camilo$^3$,
\newauthor
I.~H. Stairs$^4$, 
N. D'Amico$^{5,6}$, 
V.~M. Kaspi$^7$,
G. Hobbs$^{1,2}$,
D.~J. Morris$^1$,
F. Crawford$^8$,
\newauthor
A. Possenti$^{6,9}$,
B.~C. Joshi$^{1,10}$,
M.~A. McLaughlin$^1$,
D.~R. Lorimer$^1$,
A.~J. Faulkner$^1$
\\
$^1$ University of Manchester,
Jodrell Bank Observatory, Macclesfield, Cheshire, SK11~9DL, UK\\
$^2$ Australia Telescope National Facility, CSIRO, P.O.~Box~76, Epping
NSW~1710, Australia\\
$^3$ Columbia Astrophysics Laboratory, Columbia University, 550 West 120th
Street, New York, NY~10027, USA\\
$^4$ Department of Physics \& Astronomy, University of British Columbia,
6224 Agricultural Road, Vancouver, B.C. V6T 1Z1, Canada \\
$^5$  Dipartimento di Fisica, Universit\`a di Cagliari, 
   S.P. Monserrato-Sestu Km 0.700, I-09042 Monserrato, Italy  \\
$^6$ Osservatorio Astronomico di Cagliari, loc. Poggio dei Pini, 
   Strada 54, Capoterra, I-09012, Italy \\
$^7$ McGill University, 3600 University St., Montreal, Quebec H3T 2A8,
 Canada \\
$^8$ Department of Physics, Haverford College, Haverford, PA 19041, USA \\
$^9$ Osservatorio Astronomico di Bologna, via Ranzani 1, Bologna, 
   I-40127, Italy  \\
$^{10}$ National Center for Radio Astrophysics, P.O.~Bag 3, Ganeshkhind,
Pune, 411003 India
}

\date{2003 March 10}
\begin{document}

\maketitle
\newcommand{\setthebls}{
%                 de-comment this line for double spacing:
 %\baselineskip=20pt
}

\setthebls

\begin{abstract} 
The Parkes multibeam pulsar survey has unlocked vast areas of the
Galactic plane which were previously invisible to earlier
low-frequency and less-sensitive surveys. The survey has discovered
more than 600 new pulsars so far, including many that are young and
exotic. In this paper we report the discovery of 200 pulsars for which
we present positional and spin-down parameters, dispersion measures,
flux densities and pulse profiles. A large number of these new pulsars
are young and energetic, and we review possible associations of
$\gamma$-ray sources with the sample of about 1300 pulsars for which
timing solutions are known. Based on a statistical analysis, we
estimate that about $19\pm6$ associations are genuine.  The survey has
also discovered 12 pulsars with spin properties similar to those of
the Vela pulsar, nearly doubling the known population of such neutron
stars. Studying the properties of all known `Vela-like' pulsars, we
find their radio luminosities to be similar to normal pulsars,
implying that they are very inefficient radio sources.  Finally, we
review the use of the newly discovered pulsars as Galactic probes
and discuss the implications of the new NE2001 Galactic electron density
model for the determination of pulsar distances and luminosities.
\end{abstract}


\section{Introduction}

We increase our understanding of pulsars by studying them as a
population or by studying specific and/or exotic examples such as
binary, millisecond, young or glitching pulsars. The Parkes Multibeam
Pulsar Survey (hereafter PMPS) set out to find a large number of
pulsars previously hidden from past surveys done at low frequencies
and/or with short integration times. Distant pulsars in the Galactic
plane that were previously undetectable due to scattering and
dispersion broadening of their signals propagating through the
interstellar medium were expected to be found in large numbers.  Among
the new pulsars, a few exotic systems should be present, and the
discovery of pulsars in remote parts of the Galaxy promised to provide
a unique opportunity for population studies and the understanding of
the Galaxy.

The PMPS has fulfilled all of these promises.
It has found a large number of young 
(e.g.~Camilo et al.~2000, D'Amico et al.~2001)\nocite{ckl+00,dkm+01}, 
distant (this work), exotic
(Lyne et al.~2000, Kaspi et al.~2000b, 
Stairs et al.~2001, Camilo et al.~2001b)
\nocite{lcm+00,klm+00b,sml+01,clm+01} and glitching
pulsars (Hobbs et al.~2002)\nocite{hlj+02} buried in the inner
Galactic plane $(|b|<5^{\circ},$-100$^{\circ}<l<50^{\circ})$. 

All planned survey pointings have been completed and the first
processing of the data has yielded more than 600 newly detected
pulsars. Currently, re-processing with refined software and
acceleration search is underway (Faulkner et
al.~2002)\nocite{fkl+02}. By the time re-processing is completed, the
PMPS will have roughly doubled the number of known pulsars and opened
up vast portions of the Galaxy that were previously devoid of known
pulsars. A detailed description of the survey including the
motivation, telescope, hardware and software details, and the first
220 pulsars discovered were reported by Manchester et al.~(2001) and
Morris et al.~(2002)\nocite{mlc+01,mhl+02}, hereafter Paper I and
Paper II, respectively. Many of the more exotic pulsars have been
reported separately as noted above. All pulsars with known timing
solutions can be found in the on-line catalogue hosted by the ATNF.
\footnote{{\tt http://www.atnf.csiro.au/research/pulsar/catalogue/}}

In summary,
an operating frequency of 1374 MHz minimised the harmful effects of
dispersion and scattering that inhibit lower frequency surveys of this
region of the Galaxy. The survey has a limiting sensitivity of 
about 0.2~mJy (Paper I), 
far surpassing that of previous wide-area pulsar surveys. This
has been made possible by the 13-beam receiver system \cite{swb+96}
which enables this area of sky to be covered in a manageable time. 

In Section~\ref{sec:tim} we report the discovery of  200 new
pulsars and catalogue many of their basic parameters obtained after,
at least, one year of timing observations. We then review in
Section~\ref{sec:young} the sample of young pulsars, including the
large number discovered in this survey. We study their possible
associations with $\gamma$-ray sources (Section~\ref{sec:egret}),
using distance estimates of the new `NE2001' Galactic electron density
model \cite{cl02}. We then summarize the properties of `Vela-like'
pulsars (Section~\ref{sec:vela}), before we comment on the
implications of the NE2001 model for the Galactic distribution of
pulsars (Section~\ref{sec:gal}).

\section{Discovery and timing of 200 pulsars}
\label{sec:tim}

The observation and analysis strategies used are identical to those
outlined in Papers I and II.  Table~\ref{t:posn} lists the pulsar
name, the J2000 right ascension and declination from the timing
solution, the corresponding Galactic coordinates, the beam in which
the pulsar was detected, the radial distance of the pulsar from the
beam centre in units of the beam radius (approximately 7 arc minutes),
the signal-to-noise ratio of the discovery observation from the final
time-domain folding in the search process, the mean flux density
averaged over all observations included in the timing solution, and
pulse widths at 50\% and 10\% of the peak of the mean pulse
profile. The 10\% width is not measurable for pulsars with mean
profiles having poor signal--to-noise.  Estimated uncertainties,
where relevant, are
given in parentheses where relevant and refer to the last quoted
digit. Flux densities may be somewhat overestimated for very weak
pulsars or those which have extended null periods, since
non-detections are not included in the timing solution.
Table~\ref{t:prd} gives solar-system barycentric pulse periods, period
derivatives, epoch of the period, the number of times of arrival
(TOAs) used in the timing solution, the final rms timing residual, and
the dispersion measure.

For a few pulsars, the timing solutions include data obtained with the
76-m Lovell telescope at Jodrell Bank Observatory. Details of these
observations can be found in Paper II. Corresponding sources are
marked in Table~\ref{t:posn}. The table also includes one binary
pulsar, PSR J1141$-$6545. Binary parameters can be found in Kaspi et
al.~(2000b)\nocite{klm+00b}.  Three pulsars, PSRs J1301$-$6310,
J1702$-$4128 and J1702$-$4310, showed significant timing noise which
was removed, to first order, by fitting a second period derivative to the
data.  These pulsars are indicated in Table~\ref{t:prd}.

Table~\ref{t:deriv} lists derived parameters for the 200
pulsars. After the name, the first three columns give the base-10
logarithm of the characteristic age, $\tau_c = P/(2\dot P)$ in years,
the surface dipole magnetic field strength, $B_s = 3.2 \times 10^{19}
(P \dot P)^{1/2}$ in Gauss, and the rate of loss of rotational energy,
$\dot E = 4\pi^2 I \dot P P^{-3}$ in erg s$^{-1}$, where a
neutron-star moment of inertia $I = 10^{45}$ g cm$^2$ is assumed. The
next two columns give the pulsar distance $d$, computed from the DM
assuming the Taylor \& Cordes (1993; hereafter TC93)\nocite{tc93}
model for the Galactic distribution of free electrons, and the implied
Galactic $z$-distance.  Although distances are quoted to 0.1 kpc, in
fact they are generally more uncertain than that (typically
around 30\%) owing to
uncertainties in the electron density model. This is especially so for
pulsars with very large DMs, indicating large distances from the Sun.
Despite the availability of the improved NE2001 model, we use the TC93
model in order to be consistent with Papers I and II where the
previous model has been employed.  We discuss the uncertainties of the
TC93 distances and the difference of both models in more detail in
Section \ref{dmdiscuss}, when we compare the TC93 values to those
inferred from the new NE2001 model.

We use the TC93-model distance to compute the listed radio
luminosity, $L_{1400} \equiv S_{1400} d^2$. For a radio spectral index of
$-1.7$ \cite{mkkw00b}, these numbers may be converted to the
more commonly quoted 400 MHz luminosity by multiplying by 8.4. The
majority of all presented 
measured and derived parameters have already been included in the
statistical analyses presented in Paper II.

Mean pulse profiles at 1374 MHz for the 200 pulsars are given in
Fig.~\ref{fg:prf}. These profiles were formed by adding all data used
for the timing solution. Typically they contain several hours of effective
integration time.

\section{Young pulsars}
\label{sec:young}

One of the main aims of the PMPS
was to find young
pulsars. Here we define `young pulsars' as those with a
characteristic age less than 100 kyr.  This age is commonly
chosen as a cut-off as it includes most pulsars which are likely to
glitch often and/or to be associated with supernova remnants.

As already demonstrated in Paper II, the strategy of searching the
region close to the Galactic plane at high frequencies has indeed been
very successful in finding young pulsars: the survey has discovered 39
out of 79 currently known pulsars with $\tau_c \la 100$ kyr.

This newly increased sample of young pulsars is important in studies
of the birth properties of radio pulsars, such as the initial spin
period, their luminosity, and the kick velocity imparted by asymmetric
supernova explosions. Young pulsars also provide a possible origin for
many of the previously unidentified point sources detected by the
Energetic Gamma Ray Experiment Telescope ({\em EGRET\/}) as has been
addressed by many authors (e.g.~Merck et al.~1996).  A distinct group
of young pulsars is that of energetic objects with spin parameters
similar to Vela, which may be more typical of the young pulsar
population than the Crab pulsar which is unique in several aspects. In
the following, we will review the current sample of potential {\em
EGRET\/} counterparts and such Vela-like pulsars.

\subsection{Pulsar/{\em EGRET\/} source associations}

\label{3egsection}
\label{sec:egret}

The true nature of the 100 or so unidentified {\em EGRET\/} sources in
the Galactic plane has been debated for some time (e.g. Hartman et
al. 1999).\nocite{hbb+99} 
Pulsars are good candidates because they have a similar spatial
distribution and are one of the only two 
populations of astronomical objects positively
identified as being $\gamma$-ray emitters, as demonstrated clearly by
the Crab and the Vela pulsars.  A few recently discovered young pulsars
(see below) discovered in this survey have already been plausibly
associated with {\em EGRET\/} sources \cite{dkm+01,cbm+01,tbc01}. Here
we comment on further possible pulsar $\gamma$-rays
counterparts and discuss how
many of the positional coincidences may be genuine rather than due to
chance alignments.

In Table~\ref{t:egret} we list all pulsar/{\em EGRET\/} point source
positional associations, where the pulsar position deviates from the
nominal {\em EGRET\/} source position by less than the 95\% error box
radius, i.e.~$\Delta \Phi/\theta_{95}\le1$,
where $\Delta \Phi$ is the difference in the two positions.
We list {\em all\/} {\em
EGRET\/} point sources including those already identified as pulsars
(ID=P), galaxies (ID=G) and Active Galactic Nuclei (ID=A).  Due to the
large number of newly discovered pulsars in the Galactic plane, often
several pulsars lie in the same error box. For {\em EGRET\/} sources
that have not been identified previously (ID=?), we can attempt to
judge the likelihood of the individual possible associations by
comparing the properties of the $\gamma$-ray source and the
corresponding pulsar(s).

For the {\em EGRET\/} sources, we summarize their characteristics as
listed in the 3EG catalogue (Hartman et al.~1999). Besides the error
box size, $\theta_{95}$, we list a computed $\gamma$-ray flux,
$\bar{F}$ ($E>100$ MeV).  We have used the flux values as derived by
Hartman et al.~in units of $10^{-8}$ ph s$^{-1}$ cm$^{-2}$, assuming a
spectral index of $2.0$, to derive values in units of
erg~s$^{-1}$~cm$^{-2}$, where we followed the calculations by D'Amico
et al. (2001).\nocite{dkm+01}

It has been shown for the known genuine pulsar/{\em EGRET\/} source
associations that pulsars are steady $\gamma$-ray sources
(e.g.~McLaughlin et al.~1996).  Therefore, we also quote for all {\em
EGRET\/} sources in Table~\ref{t:egret} a variability index, $V$, as
defined by McLaughlin et al.~(1996).\nocite{mmct96} Values for $V$ are
taken from an updated list presented by
McLaughlin~(2001)\nocite{mcl01} and are typically found to be $<1$ for
pulsars.

For each {\em EGRET\/} source we list the possible pulsar counterpart(s).
Pulsar/{\em EGRET\/} source pairs that have been proposed in the literature
previously are listed with corresponding references.  Plausible
associations involving pulsars discovered in the PMPS are discussed in
more detail later. All such multibeam pulsars have been published
separately, in this work, or in Papers I or II. A few newly discovered pulsars
listed here will soon be presented with all parameters in a
forthcoming Paper IV (in preparation).

In order to assess the likelihood of an association to be genuine, one
has to compare the $\gamma$-ray luminosity, $L_\gamma$,
%as derived from the $\gamma$-ray flux, $\bar{F}$, measured by {\em EGRET\/} 
to the spin-down luminosity, $\dot{E}$, of the pulsar.  Typically,
this efficiency, $\eta \equiv L_\gamma / \dot{E}$, is 
assumed to lie in a range from
0.01\% to about 20\% (e.g.~Torres et
al.~2001\nocite{tbc01}). Deriving these numbers
requires conversion of the $\gamma$-ray flux measured by {\em EGRET},
$\bar{F}$, into the luminosity, $L_\gamma$. This is, however, a
non-trivial task, as the beaming fraction, $f$, for the high energy
emission is unknown. One typically assumes a beaming fraction of 1 sr,
i.e.~$f=1/4\pi$ (cf.~Torres et al.~2001).

In order to derive the $\gamma$-ray luminosity, one must also use a
distance estimate which is usually based on the dispersion measure.
As some pulsar distances derived from the dispersion measure differ
significantly when applying the new NE2001 electron density model
rather than the previously used TC93 model, we quote distances derived
from both models for comparison.  For instance, the distance for PSR
J0218+4232 is reduced from 5.9 kpc (TC93) to only 2.9 kpc (NE2001),
making it a more average $\gamma$-ray pulsar in terms of $\dot{E}/d^2$
and efficiency, although it remains the only millisecond pulsar
detected at $\gamma$-ray 
energies. Similarly, for PSR B1055$-$52 the new NE2001
distance estimate is much smaller than the TC93 one, resulting in a
decrease of the efficiency from 18.4\% to 4.1\%. This dramatic change
has important implications for the discussions of genuine {\em EGRET}
source/pulsar pairs as it was basically PSR B1055$-$52 which in the
past had set the upper limit for the efficiency range assumed to be
possible, i.e.~$\eta\le20$\%. The much reduced value for this pulsar
may now imply that real associations should also accommodate much
lower values for $\eta$. In the following, we will consider this
during our discussions, basing all our values such as efficiencies,
etc., on distances derived from the NE2001 model.

One can also consider the pulsar distances that would be needed for
the $\gamma$-ray luminosity to be consistent with a typically observed
efficiency, $0.01\% \la \eta_{f=1/4\pi} \la 20\%$.  Comparing these
distances to an electron density model distance is useful, as it
allows one to judge whether the uncertainties in the DM distance could
accommodate for such variations. Similarly, one can assume beaming
fractions different from $f=1/4\pi$ to further explore the possibility
of a genuine association.

Considering criteria such as source separation, variability index,
characteristic age, efficiency and distances, we finally derive a
`quality indicator', $Q$, for a proposed association.  A `+' indicates
a genuine or very likely association, while a `$-$' implies that the
apparent association is almost surely due to a chance alignment.
Pairings marked with `+?' are plausible associations, while for `?' an
association cannot be ruled out, but more sensitive instruments like
the Gamma-Ray Large Area Telescope ({\em GLAST}) are needed to study it
further.

\subsubsection{Parkes Multibeam pulsars}

The first entry of a PMPS pulsar in Table~\ref{t:egret} 
lists a positional coincidence with 3EG J1013$-$5915.  
Camilo et al.~(2001a)
argued that PSR J1016$-$5857 is a plausible counterpart to
that source and is possibly also associated with SNR G284.4$-$1.8.
Table~\ref{t:egret} also shows that there are two more
pulsars positionally coincident with the same {\em EGRET\/} source, but these
are not likely to be physically related.

Two more {\em EGRET\/} sources potentially associated with 
PMPS pulsars have already been discussed in the literature.  D'Amico et
al.~(2001)\nocite{dkm+01} 
reported the discovery of PSR J1420$-$6048 and PSR J1837$-$0604 and
discussed a possible association with
the two {\em EGRET\/} sources 3EG J1420$-$6038 and
J1837$-$0606, both of which have a variability index consistent with
known $\gamma$-ray pulsars. The first source has only one pulsar in its
error box, PSR J1420$-$6048, and is likely to be associated with it as
the pulsar's characteristic age and the derived efficiency of
$\eta_{f=1/4\pi}=1$\% are approximately as expected for a $\gamma$-ray
pulsar. Moreover, in a multi-wavelength study, Roberts et
al.~(2001\nocite{rrj01}) presented pulsed X-ray data and discussed a
possible relationship to SNR G313.6+0.3. These observations also
suggest a smaller distance than that from the TC93 model, consistent
with the NE2001 model distance.

The {\em EGRET\/} source 3EG J1837$-$0606 has two pulsars located in
its error box, but only the spin-down luminosity of PSR J1837$-$0604
presented by D'Amico et al.~(2001) is sufficient to explain the high
energy emission.  With a nominal efficiency of 7\% that is typical,
this pulsar is likely to be a genuine counterpart. The other pulsar,
J1837$-$0559, would require an unreasonably high efficiency, even when
taking uncertainties in distance and beaming fraction estimates into
account.

This paper reports the discovery of the 140-ms PMPS pulsar
J1015$-$5719. As already discussed by Torres et al.~(2001)
\nocite{tbc01} using data made available prior to publication, this
pulsar appears to be a plausible counterpart to the non-variable {\em
EGRET\/} source 3EG J1014$-$5705. The computed $\gamma$-ray efficiency
is consistent with values for $\gamma$-ray pulsars,
i.e.~$\eta_{f=1/4\pi}=7.5$\%.

The PMPS pulsar J1314$-$6101 is located on the edge of the
error box of 3EG J1308$-$6112. Its DM distance of 9.6 kpc as derived
by the Taylor \& Cordes (1993) model is reduced to 6.1 kpc in the NE2001
model.  Nevertheless, the observed spin-down luminosity is so small,
that the pulsar needs to be at a distance of only 50 pc to have a
$\gamma$-ray efficiency of even 20\%. In fact, the very large variability
index of $V=4.61$ rules out almost certainly an identification of this
{\em EGRET\/} source with a $\gamma$-ray pulsar at all.

Sturner \& Dermer (1995\nocite{sd95}) proposed that 3EG J1410$-$6147
may be associated with SNR G312.4$-$0.4. The variability index of this
{\em EGRET\/} source is consistent with observed pulsar properties,
and indeed, there are two PMPS pulsars located in the source's 95\%
error box (see Table~\ref{t:egret} and Manchester et al.~2002).  Both
pulsars have also been considered by Torres et al.~(2001) and more
recently also by Doherty et al.~(2002)\nocite{djg+02}.  PSR
J1412$-$6145 could only be consistent with the {\em EGRET\/} source if
the distance or the beaming fraction were severely overestimated. An
error in distance by a factor of 4 or so would make an association
barely acceptable, but even in the improved NE2001 model the distance
is only reduced from 9.3 to 7.8 kpc.  A more plausible counterpart
seems to be PSR J1413$-$6141 whose discovery is reported in this
paper with a  very young characteristic age of only 14 kyr.
Manchester et
al.~(2002\nocite{mbc+02}) also discussed a suggestive relationship of
this pulsar with SNR G312.4$-$0.4. Doherty et al.~(2002) recently
addressed this question in a detailed multi-wavelength study. They
argue that it was PSR J1412$-$6145 that was formed in the explosion
creating SNR G312$-$0.4, while none of the two pulsars is related to
the EGRET source.  Indeed, the NE2001 distance needs to be
overestimated, to make PSR J1413$-$6141 a plausible counterpart to 3EG
J1410$-$6147. More sensitive gamma-ray observations are needed to
settle this question.

The error box of 3EG J1639$-$4702 contains a total of five known
pulsars. The discoveries and timing parameters for three of them, PSRs
J1637$-$4642, J1637$-$4721 and J1640$-$4648, are presented in this
work.  Only PSR J1637$-$4642 has a spin-down luminosity consistent
with the {\em EGRET\/} flux density, making it a possible
counterpart. This source would have a $\gamma$-ray efficiency of about
$\eta\sim15$\% at its NE2001 distance. We note that the variability
index would be somewhat high for 3EG J1639$-$4702 to be a $\gamma$-ray
pulsar.  A fifth, relatively young PMPS pulsar has been discovered in
the same error box and appears to closest to the nominal {\em EGRET\/}
position.  Named according the current, preliminary timing solution,
PSR J1638$-$4715, however, exhibits parameters which seem to exclude
its identification as a gamma-ray pulsar.

Two new pulsars presented in this paper, PSRs J1713$-$3844 and
J1715$-$3903, both reside in the error box of 3EG J1714$-$3857. The
latter pulsar is a possible counterpart but only if its DM distance is
overestimated by a factor of 2 to 3. However, the new NE2001 distance
estimate agrees well with the previous TC93 value.  Interestingly, PSR
J1713$-$3844 is also located close to SNR G347.3$-$0.5 for which a
distance of $\sim6$ kpc is estimated \cite{sgd+99}.  This distance is
very close to the estimated pulsar distance but is too large to make
the pulsar simultaneously associated with the {\em EGRET\/}
source. The SNR harbours a compact central source detected in X-rays,
whose position is not consistent with any of the PMPS pulsars.
Recently in a targeted search Crawford et al.~(2002)\nocite{cpkm02}
discovered a weak 392-ms radio pulsar, PSR J1713$-$3949, which was
initially considered to be a possible counterpart. The current timing
solution, however, places the source just outside the {\em EGRET}
error box (i.e.~$\Delta\Phi/\theta_{95} = 1.6$) and also suggests that
this source is not consistent with the X-ray source either (Crawford,
private communication).

The error boxes of the {\em EGRET\/} sources 3EG J1638$-$5155,
J1704$-$4732, J1736$-$2908, J1741$-$2050,\\
 J1746$-$1001, J1824$-$1514, J1826$-$1302, J1837$-$0423, \\
J1837$-$0606 and J1850$-$2652 all contain
pulsars. These pulsars were either discovered in the PMPS, previously
known, or recently discovered in the Swinburne Intermediate Latitude
survey (Edwards et al.~2001)\nocite{ebvb01}. For the PMPS pulsars,
parameters have already been presented here (PSRs J1638$-$5226,
J1707$-$4729), in Paper II (PSRs J1823$-$1526, J1826$-$1526,
J1838$-$0453, J1837$-$0559, J1837$-$0604) or will be presented in a
forthcoming Paper IV (PSR J1736$-$2843, J1824$-$1505). None of the
pulsars in the error boxes of these {\em EGRET\/} sources has
parameters which suggest a physical relationship with an {\em EGRET\/}
source. A possible exception may be PSR B1823$-$13, positionally
coincident with 3EG J1826$-$1302, which would have a very reasonable
efficiency of just 2\%, but the variability index is very high.

Sturner \& Dermer (1995\nocite{sd95}) suggested that the {\em EGRET\/} source
3EG J1903+0550 is associated with SNR~G40.5$-$0.5 (W44).  Its error
box contains four pulsars, three of which are multibeam discoveries,
namely PSRs J1903$+$0609, J1905$+$0603 and J1905$+$0616.
None of the four pulsars has sufficient
spin-down luminosity to explain an association with the {\em EGRET\/} source.

\subsubsection{Pair statistics}

Table~\ref{t:egret} shows that the known population of pulsars
includes about 48 pulsars which have positional coincidence with {\em
EGRET\/} error boxes. Of these, we believe that 16 pulsar/{\em
EGRET\/} pairs are genuine or plausible associations. Note that this
number does not include Geminga and PSR B1951+32, which were not
listed in Table~\ref{t:egret}.  Geminga does not appear to be a normal
radio pulsar (Kuzmin \& Losovskii 1997, Malofeev \& Malov 1997,
McLaughlin et al.~1999), \nocite{kl97,mm97,mchm99} while PSR B1951+32
is probably detected in {\em EGRET\/} data but not associated with an
{\em EGRET\/} point source (Ramanamurthy et al.~1995).\nocite{rbd+95}
Most of the suggestive associations can only be confirmed when more
sensitive instruments like {\em GLAST} become available to detect
pulsed emission consistent with the radio pulsar period.

The number of radio pulsars detectable at high energies impacts the
understanding of the magnetospheric emission processes.  Many authors
therefore have attempted to model the small-number statistics of the genuine
{\em EGRET\/} detections (e.g.~McLaughlin \& Cordes 2000\nocite{mc00}, Zhang
et al.~2000\nocite{zzc00}, Gonthier et al.~2002\nocite{gob+02}). These
results are also used to forecast the number of $\gamma$-ray pulsars to
be detected by {\em GLAST}.  In the following we will try to answer the
question of how many pulsar/{\em EGRET\/} source pairs are likely to occur by
chance in the presently known catalogues, in a much simpler
manner. This more model-free approach follows the procedure used by Lorimer et
al.~(1998)\nocite{llc98} to study the chance alignment of
pulsar/supernova remnant pairs.

We study the normalized deviation of a pulsar from a given {\em
EGRET\/} position, $\delta\equiv\Delta\Phi/\theta_{95}$, by deriving the
distribution of $\delta$ that occurs by chance and comparing this
directly to the observed distribution. For completely unrelated sets
of pulsars and {\em EGRET\/} sources, the number of pairs occupying an
annulus between $\delta$ and $\delta+ d\delta$ is proportional to
$\delta$, regardless of the relative densities of pulsars and {\em
EGRET\/} sources over the plane of the sky (see~Lorimer et
al.~1998). One can demonstrate this by applying systematic shifts to
the positions of all known pulsars before recalculating the new
resulting distribution of $\delta$ from this shifted population which
will produce different by-chance alignments. We follow the example of
Lorimer et al.~and perform this decoupling of pulsar and {\em EGRET\/}
samples for systematic shifts of $\pm4\degr$ and $\pm8\degr$ in
Galactic longitude. Note that shifts in Galactic longitude are chosen
to avoid biases in the simulated samples due to the Galactic nature of
radio and $\gamma$-ray pulsars.  These shifts are small compared to
changes in the density of both types of objects in the sky, but much
larger than the typical error box of an {\em EGRET\/} point source.
We plot the resulting distribution of the shifted sample as well as
the observed distribution in Fig.~\ref{f:offsetplot}. The distribution
for the shifted sample shows the expected linear increase with
position deviation as indicated by the solid line.

Inspecting Fig.~\ref{f:offsetplot} we note that there is an excess of
observed pairs for $\delta\le1$. The observed pulsar sample produces
50 pairs positionally coincident with {\em EGRET\/} sources (including
now Geminga and PSR B1951+32), while the fake shifted sample produces
only $31\pm6$ pairs. Hence we have an excess of $19\pm6$ possible
associations.  This is larger than the number of fully established
associations, but consistent with our derived number of 16
associations which we classified as `+', `+?', or `?' in
Table~\ref{t:egret}.

We can consider this result as a good indication that many of the
proposed but not yet fully established associations may indeed be
real.  This view is supported by inspecting the median characteristic
age and $\dot{E}/d^2$ for pairs in a given $\delta$-interval: the
pairs for $\delta\le1$ are much younger and have much larger spin-down
fluxes (see Fig.~\ref{f:offsetplot} middle and bottom). Interestingly,
inspecting the numbers derived to produce this
Figure~\ref{f:offsetplot} this trend continues, albeit with much
smaller difference in values, even for $\delta$ somewhat larger than
unity. One can speculate as to whether this indicates a physical
relationship of {\em EGRET\/} point sources and pulsars that lie just
outside the nominal 95\% error box. Just such an example is the
aforementioned PSR B1823$-$13.  But even when excluding this source
from calculating the median $\dot{E}/d^2$, the resulting value is
still significantly larger than that for larger $\delta$'s or the
shifted sample.

The results of our simple statistical analysis agree well with
the conclusions of more complicated studies like that of McLaughlin
\& Cordes (2000)\nocite{mc00}. Based on a likelihood
analysis, they argued 
that about 20 unidentified {\em EGRET\/} sources are probably
$\gamma$-ray pulsars. This estimate of 20 sources includes radio-quiet
sources of Geminga-type, of which a second one may have been 
identified (see Halpern et al.~2002)\nocite{hgmc02}.
The similarity to our estimate of  $19\pm6$ 
genuine associations with {\em EGRET\/}
sources is therefore intriguing, even though
the involved uncertainties are considerable.

In any case, it is certain that many more $\gamma$-ray pulsars will be
detected with {\em  GLAST}, and McLaughlin \& Cordes for instance estimate up
to 750 sources.  While we can argue here only on statistical grounds,
we can nevertheless expect that a number of those new radio
pulsar/{\em GLAST} source associations will, in retrospect, also be found
in the older {\em EGRET}/pulsar data.

\subsection{Vela-like pulsars}
\label{sec:vela}

The archetypal young, energetic pulsar is PSR B0833$-$45 in the Vela
supernova remnant.  A number of other young pulsars share many
properties with Vela, such as having detectable X-ray or $\gamma$-ray
emission, being associated with pulsar wind nebulae, or being known to
exhibit instabilities in their rotation (`glitches'). The PMPS has
uncovered 12 more such Vela-like pulsars, which we will loosely define
as sources with characteristic ages in the range $10\la \tau_c \la
100$ kyr and spin-down luminosities $\dot{E}\ga 10^{36}$ erg s$^{-1}$.
The new sources almost double the number of known Vela-like pulsars,
increasing it to a total of 26. All 26 sources are listed in
Table~\ref{t:velalike} where we also quote corresponding references
and indicate which pulsars are known to glitch.  Glitch information is
obtained from the work by Shemar \& Lyne (1996), \nocite{sl96} Camilo
et al.~(2000)\nocite{ckl+00}, Wang et al.~(2000)\nocite{wmp+00} and
references therein.  As indicated, nine of these pulsars already
appeared in Table~\ref{t:egret} as genuinely or possibly related to
{\em EGRET\/} point sources.

Radio continuum maps have been obtained and/or studied for a number of
PMPS pulsars in Table~\ref{t:velalike}.  Crawford
(2000)\nocite{cra00} obtained radio maps with the Australia Telescope
Compact Array (ATCA) for PSRs J0940$-$5428, J1112$-$6103, J1301$-$6305
and J1420$-$6048. The latter pulsar (D'Amico et al.~2001 and this
work) is surrounded by a pulsar wind nebula which was studied in
detail by Roberts et al.~(1999, 2000)\nocite{rrj01}\nocite{rrjg99}.
Manchester et al.~(2002)\nocite{mbc+02} used ATCA data and maps from
the Molongo Galactic Plane Survey (Green et al.~1999)\nocite{gcly99}
to search for supernova remnants associated with PMPS pulsars.
Further results will be published elsewhere.

Using the radio
luminosity at 1400 MHz, $L_{1400}$, 
Camilo et al.~(2002b)\nocite{cmg+02} pointed out 
that young pulsars
($\tau_c\la100$kyr) are not particularly luminous by comparison with
middle aged pulsars. We can also demonstrate this for the smaller
sample of Vela-like pulsars as shown in Fig.~\ref{f:radiolum}.
While Vela-like pulsars have a median $\log L_{1400}$ of 1.5, normal
pulsars (here defined as being non-Vela-like and non-recycled
pulsars) exhibit a median of 1.4. In contrast, recycled 
or millisecond pulsars appear to be less
luminous (median 0.4), although this is to some extent due to selection
effects as discussed in detail by Kramer et al.~(1998)\nocite{kxl+98}.

We can also define an efficiency as a radio emitter by comparing 
radio luminosity to the spin-down luminosity,
$\epsilon_{1400} \equiv L_{1400}/\dot{E}$ $10^{-30}$Jy kpc$^2$ erg$^{-1}$s.
Obviously, since the the radio luminosities are
very similar for Vela-like pulsars compared to normal pulsars, while
$\dot{E}$ is much larger, it is clear that Vela-like pulsars must be
much less efficient radio pulsars. Indeed, as demonstrated in
Fig.~\ref{f:radioeff}, the medians measured for the three
distributions in $\log \epsilon_{1400}$
are $-4.8$ for Vela-like pulsars, $-1.3$ for normal pulsars and
$-3.2$ for millisecond pulsars, respectively. The result for
millisecond pulsars has been already discussed by Kramer et
al.~(1998), whilst the one for Vela-like pulsars clearly demonstrates
that spin-down and radio luminosities are not correlated for
non-recycled pulsars. It is interesting to note that when computing
the log median efficiency for energetic pulsars with
$\dot{E}\ga10^{36}$ erg s$^{-1}$ but ages larger or smaller than adopted
for Vela-like pulsars, $\log \epsilon_{1400}$
 is even lower with median $-5.8$.
However, this latter sample contains both millisecond pulsars and very young
pulsars like the Crab.

The computed radio efficiency effectively assumes that all pulsars
beam into the same solid angle. Manchester (1996) argued that young
pulsars exhibit wider beams, which may lead to an underestimation of
the efficiency for Vela-like pulsars.  We estimate that this might
account for a factor of 10 to 100, but it appears unlikely that
it explains for the full difference in the efficiency to normal
pulsars. We therefore conclude that this much enlarged sample of young
energetic Vela-like pulsars clearly demonstrates that such pulsars
lose more energy outside the radio band than normal pulsars, and that
radio luminosity does not increase with larger available spin-down
luminosity. Instead, it is possible that some saturation process is
operating.  We also note that apart from possible correlations with
age, we have also tested for possible dependences of efficiency and
luminosity on period, magnetic field at surface and light cylinder, as
well as accelerating potential above the polar caps. No correlation
has been found.

\section{Pulsars as Galactic probes}

\label{dmdiscuss}
\label{sec:gal}

Pulsars are superb objects with which to probe the Galactic structure.
In particular, the pulsars discovered in the PMPS
probe large distances and Galactic lines-of-sight 
which had largely not been accessible previously.

For the first time, the spiral arm structure becomes clearly visible
when studying the distribution of pulsars along Galactic
longitudes. In order to demonstrate this, we consider pulsars with a
characteristic age of less than 1 Myr. These pulsars are young enough
to be found close to their birthplace, even with a mean velocity of
about 450 km s$^{-1}$ (Lyne \& Lorimer 1994)\nocite{ll94}. In order to
restrict ourselves to pulsars in the Galactic disk, we only show
pulsars with Galactic latitude $|b|\le20^\circ$ in
Fig.~\ref{f:lnum}. The Galactic longitudes where our lines-of-sight
become tangents to Galactic spiral arms as given by Georgelin \&
Georgelin (1976)\nocite{gg76} (see also Cordes \& Lazio 2002) are
indicated and largely can be associated with individual peaks in the
number distributions. Interestingly, the Galactic longitude interval
$0^\circ\le l < 4^\circ$ does not contain any pulsar with determined
characteristic age of less than 1 Myr. Whilst there are newly
discovered pulsars in this interval for which the spin-down properties
still have to be determined, the known pulsars have ages larger
than 1 Myr. In contrast, there are 15 pulsars with ages less than 1 Myr in
the interval $356^\circ\le l < 360^\circ$. The general dip in the
distribution found around the Galactic Centre can in part be
attributed to selection effects, i.e.~enhanced scatter broadening,
preventing the discovery of fast rotating pulsars in the innermost
Galaxy for frequencies below about 5 GHz (see Cordes \& Lazio 1997,
Kramer et al.~2000)\nocite{cl97,kkl+00}.  On-going population
synthesis studies will investigate this effect further, and results
will be presented elsewhere.

We can expect that this structure in the number density
should also be reflected in the observed dispersion measure
distribution. In Fig.~\ref{f:maxmeandm} we show the maximum
and mean dispersion measure of all pulsars with $|b|\le20^\circ$.
Indeed, individual spiral arms can be easily identified.

The dispersion measures of the newly discovered pulsars, as well as
future measurements of scatter broadening times, obviously provide
extremely valuable input to any modelling of the Galactic free electron
density distribution. Despite the survey's limitation to Galactic
latitudes of $|b|\le5^\circ$, it also contributes
significantly to studies of the scale-height of the electron density
above the Galactic plane, as we demonstrate in Fig.~\ref{f:bdm}.

In the simplest model, we can describe the free electron
distribution in a thin slab model, with a constant electron density,
$n_e$, and a height of $\pm H$ above and below the Galactic plane. It
is then easy to show that the maximum possible dispersion measure
along a Galactic latitude $b$, is given by DM$_{max}=n_e \times H/\sin
b$. The envelope describing the data shown in Fig.~\ref{f:bdm} is
given by DM$_{max}=18/\sin|b|$ cm$^{-3}$pc, hence, $n_e\times H \sim
18$ cm$^{-3}$pc. With a canonical electron density of $n_e \sim 0.03$
cm$^{-3}$, we obtain a scale-height of 600 pc. Obviously, in realistic
models the electron density depends on Galactocentric radius and
$z$-height above and below the Galactic plane, This is the case for
both the TC93 and NE2001 models, which use an ``outer thick
component'' deriving scale-heights of about 1 kpc.

Besides revealing the large-scale structure of the Milky Way, models
of the free electron distribution are important to determine distance
estimates of pulsars. A distance is derived by integrating the
electron density in a given model along the line-of-sight towards the
pulsar until the dispersion measure is reached. Such dispersion
measure distances based on the TC93 electron density model are quoted
in Table~\ref{t:deriv}, consistent with Paper I and II. They were
particularly important in Section~\ref{3egsection}, where we also used
distances derived from the new NE2001 model as shown in
Table~\ref{t:egret}. A reliable conversion from dispersion measure to
distance and vice versa is therefore highly desirable.

The NE2001 model already incorporates a large number of dispersion
measures from multibeam pulsars, so that with the new model fitting
procedure developed by Cordes \& Lazio (2002), we can expect a
significant improvement of the model, in particular for distant
pulsars.  This is indeed the case, as we demonstrate in the following.
The location of all known pulsars in the Galactic plane as derived
from the TC93 model is shown in Fig.~\ref{f:planetc93} whereas we
applied the NE2001 model in Fig.~\ref{f:planene2001}. Newly discovered
pulsars in the PMPS are marked as open diamonds. Once more, only
pulsars with a Galactic latitude $|b|\le20^\circ$ are shown.  The
spiral arms in these figures are those used by Cordes \& Lazio (2002)
and are based on the model by Georgelin \& Georgelin
(1976)\nocite{gg76} which was also used by Taylor \& Cordes (1993). We
point out that the apparent location of most pulsars along spiral arms
has to be viewed with care since both TC93 and NE2001 incorporate
explicitly this model of the spiral arm structure of the Galaxy,
shapes and locations of which are derived from radio and optical
observations.  However, as pointed out by Taylor \& Cordes (1993), and
Cordes \& Lazio (2002) and is clear from Figure~\ref{f:lnum}, the
data make an inclusion of a spiral structure mandatory.

Limitations of the TC93 model are immediately visible from
Fig.~\ref{f:planetc93} since a number of pulsars are located far
outside the Galaxy, in particular along a semi-circle with radius of
30 kpc around the Sun.  This artifact is caused by terminating the
integration along the line-of-sight at that distance. The electron
density is obviously underestimated towards these directions in the
TC93 model. In contrast, the NE2001 model improves on the distribution
significantly, also causing the 30-kpc circle to disappear. This goes
along with a general decrease of pulsar distances, sometimes to a
large extent as mentioned in Section~\ref{3egsection}.

A similar effect is seen in changes for the computed $z$-height above
or below the Galactic plane. Figures~\ref{f:zlold} and \ref{f:zlnew}
shows the magnitude of $z$-height computed from the TC93 and NE2001
models, respectively, as a function of derived DM distance. Again, we
restrict the sample shown to pulsars with $|b|\le20^\circ$.  As
before, the TC93 model runs out of electrons before the integration
stops, producing artifacts in the resulting distribution.  In stark
contrast, the NE2001 model pulls the pulsars much closer towards the
Sun and therefore also closer to the plane. There seems to be a
paucity of pulsars in a region of large distances and large
z-heights.  We consider it unlikely that this can be attributed to
a simple selection effect. In both cases, the vast majority of distant
pulsars has been found in the PMPS. In spite searching only
latitudes of $|b|\le5^\circ$, this corresponds to a $|z|$-height of 1.3
kpc in 15 kpc distance and 2.6 kpc in 30 kpc distance, hence covering
this area in principle.

A viable test for every electron density model is to check the
existence of an (artificial) dependence of the computed $z$-height on
the estimated distance. We make this test by computing the median of
the absolute $z$-height in 2-kpc intervals for distances below 10 kpc,
and in 10-kpc intervals for distances beyond. These medians are shown
as filled diamonds in Figures~\ref{f:zlold} and \ref{f:zlnew},
centered on the corresponding intervals. Note that we choose the
median rather than the mean to account for the demonstrated artifacts
produced by the TC93 model.

The values for the TC93 model remain essentially constant up to 10
kpc.  Beyond that distance, the small-number statistics results in
large fluctuations with an apparent increase in the medians.  In the
NE2001 model, the pulsars are significantly closer to the disk, and a
slight trend is visible in the medians to decrease with distance.

In summary, the PMPS pulsars presented here and the associated
papers provide an excellent tool to study the structure of the Milky
Way. The improved understanding of this structure feeds back into our
understanding of pulsars. When, for instance, the distance of pulsars
located in far Galactic regions identified above become more
reliable, we can learn more about the pulsars' luminosity, their
distribution in the Milky Way and ultimately their population as a
whole.

\section{Summary \& Conclusions}

We have presented the parameters and pulse profiles for 200 pulsars
newly discovered in the Parkes multibeam pulsar survey. We paid
particular attention to young pulsars for which we review the
situation of possible associations with {\em EGRET\/} point
sources. In a statistical analysis we showed that a number of new
associations emerging from the Parkes multibeam pulsar sample are
likely to be genuine. We summarized the properties of the sample of
Vela-like pulsars, of which many new examples are presented in this
paper. We found that Vela-like pulsars are less efficient radio
emitters than normal pulsars as their radio luminosity does not scale
with the available spin-down luminosity.  Finally, we demonstrated
that the many new discoveries of distant pulsars in the Galactic plane
help to significantly improve the model of the free electron
distribution. For the first time, the spiral structure of the
Galaxy is directly visible in the number distribution of pulsars
along Galactic longitude.

In coming years we can expect further follow-up studies and investigations
based on this unique sample of new pulsars.

\section*{Acknowledgements} 
We gratefully acknowledge the technical assistance with hardware and
software provided by Jodrell Bank Observatory, CSIRO ATNF,
Osservatorio Astronomico de Bologna, and Swinburne Centre for
Astrophysics and Super-computing.  The Parkes radio telescope is part
of the Australia Telescope which is funded by the Commonwealth of
Australia for operation as a National Facility managed by CSIRO.  FC
is supported by NASA grant NAG~5-9950. DRL is a University Research
Fellow funded by the Royal Society.  MAM is a NSF MPS-DRF Fellow. VMK
is a Canada Research Chair and is supported NSERC, NATEQ and CIAR. IHS
holds an NSERC University Faculty Award and is supported by a Discovery
Grant.

%%%%%%%%%%%%%%%%%%%%%%%%%%%%%
%\end{document}
%%%%%%%%%%%%%%%%%%%%%%%%%%%%%

%\bibliographystyle{mn}
%\bibliography{journals,mod_jfb,modrefs,psrrefs,crossrefs}
%\bibliography{journals,modrefs,psrrefs,crossrefs}

\begin{thebibliography}{{{Roberts}, {Romani} \& {Johnston} }{2001}}

\bibitem[\protect\citename{Camilo {\rm et~al. }}{2000}]{ckl+00}
Camilo~F., Kaspi~V.~M., Lyne~A.~G., Manchester~R.~N., Bell~J.~F., D'Amico~N.,
  McKay~N. P.~F., Crawford~F., 2000, ApJ, 541, 367

\bibitem[\protect\citename{Camilo {\rm et~al. }}{2001a}]{cbm+01}
Camilo~F. {\rm et~al.}, 2001a, ApJ, 557, L51

\bibitem[\protect\citename{Camilo {\rm et~al. }}{2001b}]{clm+01}
Camilo~F. {\rm et~al.}, 2001b, ApJ, 548, L187

\bibitem[\protect\citename{Camilo {\rm et~al. }}{2002a}]{cmgl02}
Camilo~F., Manchester~R.~N., Gaensler~B.~M., Lorimer~D.~R., 2002a, ApJ,
579, L25

\bibitem[\protect\citename{{Camilo} {\rm et~al. }}{2002b}]{cmg+02}
{Camilo}~F., {Manchester}~R.~N., {Gaensler}~B.~M., {Lorimer}~D.~R.,
  {Sarkissian}~J., 2002b, ApJ, 567, L71

\bibitem[\protect\citename{{Camilo} {\rm et~al. }}{2002c}]{csl+02}
{Camilo}~F. {\rm et~al.}, 2002c, ApJ, 571, L41

%\bibitem[\protect\citename{{Caswell} \& {Barnes} }{1985}]{cb85}
%{Caswell}~J.~L., {Barnes}~P.~J., 1985, MNRAS, 216, 753

\bibitem[\protect\citename{Clifton {\rm et~al. }}{1992}]{clj+92}
Clifton~T.~R., Lyne~A.~G., Jones~A.~W., McKenna~J., Ashworth~M., 1992, MNRAS,
  254, 177

\bibitem[\protect\citename{Cordes \& Lazio }{1997}]{cl97}
Cordes~J.~M., Lazio~J. T.~W., 1997, ApJ, 475, 557

\bibitem[\protect\citename{{Cordes} \& {Lazio} }{2002}]{cl02}
{Cordes}~J.~M., {Lazio}~T.~J.~W., 2002, preprint, astro-ph/0207156

\bibitem[\protect\citename{Crawford }{2000}]{cra00}
Crawford~F., 2000, {\rm PhD thesis}, Massachusetts Institute of Technology

\bibitem[\protect\citename{{Crawford} {\rm et~al. }}{2001}]{ckm+01}
{Crawford}~F., {Kaspi}~V.~M., {Manchester}~R.~N., {Lyne}~A.~G., {Camilo}~F.,
  {D'Amico}~N., 2001, ApJ, 553, 367

\bibitem[\protect\citename{{Crawford} {\rm et~al. }}{2002}]{cpkm02}
{Crawford}~F., {Pivovaroff}~M.~J., {Kaspi}~V.~M., {Manchester}~R.~N., 2002, in
  ASP Conf. Ser. 271: Neutron Stars in Supernova Remnants., 
eds.~Patrick O. Slane and Bryan M. Gaensler, San Francisco: ASP,
\newblock p.~37

\bibitem[\protect\citename{D'Amico {\rm et~al. }}{1998}]{dsb+98}
D'Amico~N., Stappers~B.~W., Bailes~M., Martin~C.~E., Bell~J.~F., Lyne~A.~G.,
  Manchester~R.~N., 1998, MNRAS, 297, 28

\bibitem[\protect\citename{D'Amico {\rm et~al. }}{2001}]{dkm+01}
D'Amico~N. {\rm et~al.}, 2001, ApJ, 552, L45

\bibitem[\protect\citename{Doherty {\rm et~al. }}{2002}]{djg+02}
Doherty~M., Johnston~S., Green~A.~J., Roberts~M.~S.~E., Romani~R.~W.,
Gaensler~B.M., Crawford~F., 2003, MNRAS, 339, 1048

\bibitem[\protect\citename{Edwards \& Bailes }{2001}]{eb01b}
Edwards~R.~T., Bailes~M., 2001, ApJ, 801, 553

\bibitem[\protect\citename{{Edwards} {\rm et~al. }}{2001}]{ebvb01}
{Edwards}~R.~T., {Bailes}~M., {van Straten}~W., {Britton}~M.~C., 2001, MNRAS,
  326, 358


\bibitem[\protect\citename{Faulkner {\rm et~al. }}{2002}]{fkl+02}
Faulkner~A. {\rm et~al.}, 2003, 
in Radio Pulsars, eds.~M.~Bailes, D.~Nice, S.~Thorsett, M.~Bailes, 
ASP Conference Series, in press

\bibitem[\protect\citename{Fierro {\rm et~al. }}{1993}]{fbb+93}
Fierro~J.~M. {\rm et~al.}, 1993, ApJ, 413, L27

\bibitem[\protect\citename{Georgelin \& Georgelin }{1976}]{gg76}
Georgelin~Y.~M., Georgelin~Y.~P., 1976, A\&A, 49, 57

\bibitem[\protect\citename{{Gonthier} {\rm et~al. }}{2002}]{gob+02}
{Gonthier}~P.~L., {Ouellette}~M.~S., {Berrier}~J., {O'Brien}~S.,
  {Harding}~A.~K., 2002, ApJ, 565, 482

\bibitem[\protect\citename{Green {\rm et~al. }}{1999}]{gcly99}
Green~A.~J., Cram~L.~E., Large~M.~I., Ye~T., 1999, ApJS, 122, 207

\bibitem[\protect\citename{{Halpern} {\rm et~al. }}{2001}]{hcg+01}
{Halpern}~J.~P., {Camilo}~F., {Gotthelf}~E.~V., {Helfand}~D.~J., {Kramer}~M.,
  {Lyne}~A.~G., {Leighly}~K.~M., {Eracleous}~M., 2001, ApJ, 552, L125

\bibitem[\protect\citename{{Halpern} {\rm et~al. }}{2002}]{hgmc02}
{Halpern}~J.~P., {Gotthelf}~E.~V., {Mirabal}~N., {Camilo}~F., 
2002, ApJ, 573, L41


\bibitem[\protect\citename{Hartman {\rm et~al. }}{1999}]{hbb+99}
Hartman~R.~C. {\rm et~al.}, 1999, ApJS, 123, 79

\bibitem[\protect\citename{Hobbs {\rm et~al. }}{2002}]{hlj+02}
Hobbs~G. {\rm et~al.}, 2002, MNRAS, 333, L7

\bibitem[\protect\citename{Hulse \& Taylor }{1974}]{ht74}
Hulse~R.~A., Taylor~J.~H., 1974, ApJ, 191, L59

\bibitem[\protect\citename{Johnston {\rm et~al. }}{1992}]{jml+92}
Johnston~S., Manchester~R.~N., Lyne~A.~G., Bailes~M., Kaspi~V.~M., Qiao~G.,
  D'Amico~N., 1992, ApJ, 387, L37

\bibitem[\protect\citename{Kaspi {\rm et~al. }}{1996}]{kmj+96}
Kaspi~V.~M., Manchester~R.~N., Johnston~S., Lyne~A.~G., D'Amico~N., 1996, AJ,
  111, 2028

\bibitem[\protect\citename{Kaspi {\rm et~al. }}{1997}]{kbm+97}
Kaspi~V.~M., Bailes~M., Manchester~R.~N., Stappers~B.~W., Sandhu~J.~S.,
  Navarro~J., D'Amico~N., 1997, ApJ, 485, 820

\bibitem[\protect\citename{Kaspi {\rm et~al. }}{2000}]{klm+00}
Kaspi~V.~M., Lackey~J.~R., Mattox~J., Manchester~R.~N., Bailes~M., Pace~R.,
  2000a, ApJ, 528, 445

\bibitem[\protect\citename{Kaspi {\rm et~al. }}{2000}]{klm+00b}
Kaspi~V.~M., Lyne~A.~G., Manchester~R.~N., Crawford~F., Camilo~F.,
Bell~J.~F., D'Amico~N., Stairs,~I.~H., McKay~N.~P.~F., Morris~D.~J.,
Possenti~A.,, 2000b, ApJ, 543, 321

\bibitem[\protect\citename{Kramer {\rm et~al. }}{1998}]{kxl+98}
Kramer~M., Xilouris~K.~M., Lorimer~D.~R., Doroshenko~O., Jessner~A.,
  Wielebinski~R., Wolszczan~A., Camilo~F., 1998, ApJ, 501, 270

\bibitem[\protect\citename{Kramer {\rm et~al. }}{2000}]{kkl+00}
Kramer~M., Klein~B., Lorimer~D.~R., M\"uller~P., Jessner~A., Wielebinski~R.,
  2000, in Kramer~M., Wex~N., Wielebinski~R., eds, Pulsar Astronomy - 2000 and
  Beyond, {IAU} Colloquium 177.
\newblock Astronomical Society of the Pacific, San Francisco, p.~37

\bibitem[\protect\citename{Kuiper {\rm et~al. }}{2000}]{khv+00}
Kuiper~L., Hermsen~W., Verbunt~F., Lyne~A.~G., Stairs~I.~H., Thompson~D.~J.,
  Cusumano~G., 2000, in Kramer~M., Wex~N., Wielebinski~R., eds, Pulsar
  Astronomy - 2000 and Beyond, {IAU} Colloquium 177.
\newblock Astronomical Society of the Pacific, San Francisco, p.~355

\bibitem[\protect\citename{Kulkarni {\rm et~al. }}{1988}]{kcb+88}
Kulkarni~S.~R., Clifton~T.~R., Backer~D.~C., Foster~R.~S., Fruchter~A.~S.,
  Taylor~J.~H., 1988, Nature, 331, 50

\bibitem[\protect\citename{Kuzmin \& Losovskii }{1997}]{kl97}
Kuzmin~A.~D., Losovskii~B.~Y., 1997, 23, 283

\bibitem[\protect\citename{Large, Vaughan \& Mills }{1968}]{lvm68}
Large~M.~I., Vaughan~A.~E., Mills~B.~Y., 1968, Nature, 220, 340

\bibitem[\protect\citename{Lorimer, Lyne \& Camilo }{1998}]{llc98}
Lorimer~D.~R., Lyne~A.~G., Camilo~F., 1998, A\&A, 331, 1002

\bibitem[\protect\citename{Lyne \& Lorimer }{1994}]{ll94}
Lyne~A.~G., Lorimer~D.~R., 1994, Nature, 369, 127

\bibitem[\protect\citename{Lyne {\rm et~al. }}{2000}]{lcm+00}
Lyne~A.~G. {\rm et~al.}, 2000, MNRAS, 312, 698

\bibitem[\protect\citename{Malofeev \& Malov }{1997}]{mm97}
Malofeev~V.~M., Malov~O.~I., 1997, Nature, 389, 697

\bibitem[\protect\citename{{Manchester}}{1996}]{man96}
{Manchester}~R.~N., 1996, in ASP Conf. Ser. 105: 
{Pulsars: Problems and Progress, {IAU} Colloquium 160},
eds.~{S. Johnston and M.~A. Walker and M. Bailes},
San Francisco: ASP,
\newblock p.~193


\bibitem[\protect\citename{Manchester, D'Amico \& Tuohy }{1985}]{mdt85}
Manchester~R.~N., D'Amico~N., Tuohy~I.~R., 1985, MNRAS, 212, 975

\bibitem[\protect\citename{Manchester {\rm et~al. }}{1978}]{mlt+78}
Manchester~R.~N., Lyne~A.~G., Taylor~J.~H., Durdin~J.~M., Large~M.~I.,
  Little~A.~G., 1978, MNRAS, 185, 409

\bibitem[\protect\citename{Manchester {\rm et~al. }}{1993}]{mml+93}
Manchester~R.~N., Mar~D., Lyne~A.~G., Kaspi~V.~M., Johnston~S., 1993, ApJ, 403,
  L29

\bibitem[\protect\citename{Manchester {\rm et~al. }}{2001}]{mlc+01}
Manchester~R.~N. {\rm et~al.}, 2001, MNRAS, 328, 17

\bibitem[\protect\citename{{Manchester} {\rm et~al. }}{2002}]{mbc+02}
{Manchester}~R.~N. {\rm et~al.}, 2002, in ASP Conf. Ser. 271: Neutron Stars in
  Supernova Remnants.
eds.~Patrick O. Slane and Bryan M. Gaensler, San Francisco: ASP,
\newblock p.~31

\bibitem[\protect\citename{{Maron} {\rm et~al. }}{2000}]{mkkw00b}
{Maron}~O., {Kijak}~J., {Kramer}~M., {Wielebinski}~R., 2000, A\&AS, 147, 195

\bibitem[\protect\citename{{McLaughlin} \& {Cordes} }{2000}]{mc00}
{McLaughlin}~M.~A., {Cordes}~J.~M., 2000, ApJ, 538, 818

\bibitem[\protect\citename{McLaughlin }{2001}]{mcl01}
McLaughlin~M.~A., 2001, {\rm PhD thesis}, Cornell University

\bibitem[\protect\citename{McLaughlin {\rm et~al. }}{1996}]{mmct96}
McLaughlin~M.~A., Mattox~J.~R., Cordes~J.~M., Thompson~D.~J., 1996, ApJ, 473,
  763

\bibitem[\protect\citename{{McLaughlin} {\rm et~al. }}{1999}]{mchm99}
{McLaughlin}~M.~A., {Cordes}~J.~M., {Hankins}~T.~H., {Moffett}~D.~A., 1999,
  ApJ, 512, 929

\bibitem[\protect\citename{Merck {\rm et~al. }}{1996}]{mbd+96}
Merck~M. {\rm et~al.}, 1996, A\&AS, 120, 465

\bibitem[\protect\citename{{Morris} {\rm et~al. }}{2002}]{mhl+02}
{Morris}~D.~J. {\rm et~al.}, 2002, MNRAS, 335, 275

\bibitem[\protect\citename{Ramanamurthy {\rm et~al. }}{1995}]{rbd+95}
Ramanamurthy~P.~V. {\rm et~al.}, 1995, ApJ, 447, L109

\bibitem[\protect\citename{Roberts {\rm et~al. }}{1999}]{rrjg99}
Roberts~M. S.~E., Romani~R.~W., Johnston~S., Green~A.~J., 1999, ApJ, 515, 712

\bibitem[\protect\citename{{Roberts} {\rm et~al. }}{2002}]{rhr+02}
{Roberts}~M.~S.~E., {Hessels}~J.~W.~T., {Ransom}~S.~M., {Kaspi}~V.~M.,
  {Freire}~P.~C.~C., {Crawford}~F., {Lorimer}~D.~R., 2002, ApJ, 577, L19

\bibitem[\protect\citename{{Roberts}, {Romani} \& {Johnston} }{2001}]{rrj01}
{Roberts}~M.~S.~E., {Romani}~R.~W., {Johnston}~S., 2001, ApJ, 561, L187

\bibitem[\protect\citename{Shemar \& Lyne }{1996}]{sl96}
Shemar~S.~L., Lyne~A.~G., 1996, MNRAS, 282, 677

\bibitem[\protect\citename{{Slane} {\rm et~al. }}{1999}]{sgd+99}
{Slane}~P., {Gaensler}~B.~M., {Dame}~T.~M., {Hughes}~J.~P., {Plucinsky}~P.~P.,
  {Green}~A., 1999, ApJ, 525, 357

\bibitem[\protect\citename{Staelin \& Reifenstein }{1968}]{sr68}
Staelin~D.~H., Reifenstein~{III}~E.~C., 1968, Science, 162, 1481

\bibitem[\protect\citename{Stairs {\rm et~al. }}{2001}]{sml+01}
Stairs~I.~H. {\rm et~al.}, 2001, MNRAS, 325, 979

\bibitem[\protect\citename{Staveley-Smith {\rm et~al. }}{1996}]{swb+96}
Staveley-Smith~L. {\rm et~al.}, 1996, PASA, 13, 243

\bibitem[\protect\citename{Sturner \& Dermer }{1995}]{sd95}
Sturner~S.~J., Dermer~C.~D., 1995, A\&A, 293, L17

\bibitem[\protect\citename{Taylor \& Cordes }{1993}]{tc93}
Taylor~J.~H., Cordes~J.~M., 1993, ApJ, 411, 674

\bibitem[\protect\citename{Thompson {\rm et~al. }}{1992}]{tab+92}
Thompson~D.~J. {\rm et~al.}, 1992, Nature, 359, 615

\bibitem[\protect\citename{{Torres}, {Butt} \& {Camilo} }{2001}]{tbc01}
{Torres}~D.~F., {Butt}~Y.~M., {Camilo}~F., 2001, ApJ, 560, L155

\bibitem[\protect\citename{Toscano {\rm et~al. }}{1998}]{tbms98}
Toscano~M., Bailes~M., Manchester~R., Sandhu~J., 1998, ApJ, 506, 863

\bibitem[\protect\citename{Wang {\rm et~al. }}{2000}]{wmp+00}
Wang~N., Manchester~R.~N., Pace~R., Bailes~M., Kaspi~V.~M., Stappers~B.~W.,
  Lyne~A.~G., 2000, MNRAS, 317, 843

\bibitem[\protect\citename{Wolszczan, Cordes \& Dewey }{1991}]{wcd91}
Wolszczan~A., Cordes~J.~M., Dewey~R.~J., 1991, ApJ, 372, L99

\bibitem[\protect\citename{{Zhang}, {Zhang} \& {Cheng} }{2000}]{zzc00}
{Zhang}~L., {Zhang}~Y.~J., {Cheng}~K.~S., 2000, A\&A, 357, 957

\end{thebibliography}

%%%%%%%%%%%%%%%%%%%%%%%%%%%%%%%%%%%%%%%%%%%%%%%%%%%
\clearpage
\begin{table*}
\begin{minipage}{150mm}
\caption{\label{t:posn}
Positions, flux densities and widths for 200 pulsars discovered in
Parkes multibeam pulsar survey.
All pulsars were timed using the Parkes telescope. 
'J' indicates pulsars that have also been timed at Jodrell Bank.
Radial angular distances are given in units of beam radii.
}
\begin{tabular}{lllrrccrllr}
\hline
\multicolumn{1}{c}{PSR J} & R.A. (J2000) & Dec. (J2000) & 
\multicolumn{1}{c}{$l$} & \multicolumn{1}{c}{$b$} & Beam & Radial &  
 \multicolumn{1}{c}{$S/N$} & \multicolumn{1}{c}{$S_{1400}$} & \multicolumn{1}{c}{$W_{50}$} & 
\multicolumn{1}{c}{$W_{10}$} \\
  & (h~~~m~~~s) & (~\degr ~~~\arcmin ~~~\arcsec) & 
\multicolumn{1}{c}{(\degr)} & \multicolumn{1}{c}{(\degr)} &   & Dist. &  &
\multicolumn{1}{c}{(mJy)} & \multicolumn{1}{c}{(ms)} & \multicolumn{1}{c}{(ms)} \\ 
\hline
0831$-$4406  & 08:31:32.43(3)  & $-$44:06:11.9(4)  & 262.29 & $-$2.69 & !6 & 0.85 &   54.9 &   0.43(5) &  !!6.1 &   17 \\
0834$-$4159  & 08:34:16.3(1)   & $-$41:59:51(1)    & 260.89 & $-$1.04 & !3 & 0.43 &   25.6 &   0.19(3) &  !!5.8 &  -- \\
0855$-$4644  & 08:55:36.18(3)  & $-$46:44:13.4(5)  & 266.97 & $-$1.00 & !3 & 0.25 &   15.6 &   0.20(3) &  !!7.5 &  -- \\
0855$-$4658  & 08:55:19.57(5)  & $-$46:58:22.6(9)  & 267.12 & $-$1.19 & !7 & 0.19 &   39.7 &   0.23(3) &  !11.0 &   22 \\
0945$-$4833  & 09:45:38.254(6) & $-$48:33:14.50(5) & 274.20 & +3.67 & !5 & 0.51 &   39.5 &   0.39(5) &  !!4.6 &   11 \\
\\
\hline
\end{tabular}   
\label{tb:posn}
\end{minipage}
\end{table*}

\clearpage
\addtocounter{table}{-1}
\begin{table*}
\begin{minipage}{150mm}
\caption{-- {\it continued}}
\begin{tabular}{lllrrccrllr}
\hline
\multicolumn{1}{c}{PSR J} & R.A. (J2000) & Dec. (J2000) & 
\multicolumn{1}{c}{$l$} & \multicolumn{1}{c}{$b$} & Beam & Radial &
\multicolumn{1}{c}{$S/N$} & \multicolumn{1}{c}{$S_{1400}$} & \multicolumn{1}{c}{$W_{50}$} & 
\multicolumn{1}{c}{$W_{10}$} \\
  & (h~~~m~~~s) & (~\degr ~~~\arcmin ~~~\arcsec) & 
\multicolumn{1}{c}{(\degr)} & \multicolumn{1}{c}{(\degr)} &   & Dist. &  &
\multicolumn{1}{c}{(mJy)} & \multicolumn{1}{c}{(ms)} & \multicolumn{1}{c}{(ms)} \\ 
\hline
1306$-$6242  & 13:06:44.6(1)   & $-$62:42:03(1)    & 304.69 & +0.12 & !7 & 1.02 &    9.1 & 0.14(2) &    !33 &    -- \\
1309$-$6526  & 13:09:00.29(8)  & $-$65:26:16.6(5)  & 304.76 & $-$2.63 & !5 & 0.64 &   11.6 &   0.15(3) &  !10.0 &    -- \\
1314$-$6101  & 13:14:23.4(9)   & $-$61:01:16(6)    & 305.71 & +1.73 & !2 & 0.93 &   46.2 &   0.41(5) &    !59 &   86 \\
1319$-$6105  & 13:19:26.32(2)  & $-$61:05:26.2(1)  & 306.31 & +1.60 & !1 & 1.78 &   16.1 &   0.84(9) &  !13.0 &   26 \\
1322$-$6329  & 13:22:18.0(2)   & $-$63:29:37(2)    & 306.37 & $-$0.83 & !7 & 0.39 &   13.3 &   0.17(3) &    !34 &    -- \\
\\
\hline
\end{tabular}   
\end{minipage}
\end{table*}

\clearpage
\addtocounter{table}{-1}
\begin{table*}
\begin{minipage}{150mm}
\caption{-- {\it continued}}
\begin{tabular}{lllrrccrllr}
\hline
\multicolumn{1}{c}{PSR J} & R.A. (J2000) & Dec. (J2000) & 
\multicolumn{1}{c}{$l$} & \multicolumn{1}{c}{$b$} & Beam & Radial &
\multicolumn{1}{c}{$S/N$} & \multicolumn{1}{c}{$S_{1400}$} & \multicolumn{1}{c}{$W_{50}$} & 
\multicolumn{1}{c}{$W_{10}$} \\
  & (h~~~m~~~s) & (~\degr ~~~\arcmin ~~~\arcsec) & 
\multicolumn{1}{c}{(\degr)} & \multicolumn{1}{c}{(\degr)} &   & Dist. &  &
\multicolumn{1}{c}{(mJy)} & \multicolumn{1}{c}{(ms)} & \multicolumn{1}{c}{(ms)} \\ 
\hline
1535$-$5848  & 15:35:16.76(4)  & $-$58:48:27.7(3)  & 322.80 & $-$2.41 & !3 & 0.93 &   21.6 &   0.35(4) &  !!5.8 &   12 \\
1537$-$5153  & 15:37:15.73(6)  & $-$51:53:06(1)    & 327.09 & +3.04 & 10 & 0.92 &   12.7 & 0.07(2) &  !23 &   -- \\
1538$-$5638  & 15:38:05.7(2)   & $-$56:38:12(4)    & 324.38 & $-$0.87 & 10 & 1.16 &   14.0 &   0.28(4) &    !46 &    -- \\
1538$-$5750  & 15:38:08.41(4)     & $-$57:50:17.1(6)    & 323.72 & $-$1.83 & !6 & 0.07 &   13.4 & 0.06(2) &  !!4.7 &  145 \\
1539$-$5521  & 15:39:07.97(7)  & $-$55:21:11.2(8)  & 325.26 & +0.08 & !2 & 0.57 &    8.9 & 0.14(2) &    !23 &    -- \\
\\
\hline
\end{tabular}   
\end{minipage}
\end{table*} 

\clearpage
\addtocounter{table}{-1}
\begin{table*}
\begin{minipage}{150mm}
\caption{-- {\it continued}}
\begin{tabular}{lllrrccrllr}
\hline
\multicolumn{1}{c}{PSR J} & R.A. (J2000) & Dec. (J2000) & 
\multicolumn{1}{c}{$l$} & \multicolumn{1}{c}{$b$} & Beam & Radial &
\multicolumn{1}{c}{$S/N$} & \multicolumn{1}{c}{$S_{1400}$} & \multicolumn{1}{c}{$W_{50}$} & 
\multicolumn{1}{c}{$W_{10}$} \\
  & (h~~~m~~~s) & (~\degr ~~~\arcmin ~~~\arcsec) & 
\multicolumn{1}{c}{(\degr)} & \multicolumn{1}{c}{(\degr)} &   & Dist. &  &
\multicolumn{1}{c}{(mJy)} & \multicolumn{1}{c}{(ms)} & \multicolumn{1}{c}{(ms)} \\ 
\hline
1649$-$4653  & 16:49:24.5(1)   & $-$46:53:09(6)    & 339.02 & $-$1.38 & 10 & 0.80 &   16.0 &   0.31(4) &  !18.0 &    -- \\
1650$-$4126  & 16:50:13.174(5) & $-$41:26:33.8(3)  & 343.29 & +2.00 & 13 & 1.02 &   14.8 &   0.29(4) &  !!7.6 &   15 \\
1650$-$4341  & 16:50:44.68(1)  & $-$43:41:30.8(5)  & 341.62 & +0.48 & !2 & 0.64 &   14.8 &   0.26(4) &    !20 &    -- \\
1651$-$4519  & 16:51:57.2(5)   & $-$45:19:11(8)    & 340.51 & $-$0.72 & !3 & 1.06 &   18.5 &   0.54(6) &    !52 &    -- \\
1653$-$4315  & 16:53:29.7(2)   & $-$43:15:01(5)    & 342.28 & +0.38 & 10 & 0.39 &   21.8 &   0.53(6) &   !105 &    -- \\
\\
\hline
\end{tabular}   
\end{minipage}
\end{table*}

\clearpage
\begin{table*}
\begin{minipage}{150mm}
\caption{\label{t:prd}
Periods, period derivatives and dispersion measures for 200 
pulsars discovered in Parkes multibeam pulsar survey.
We also give the MJD of the epoch used for
period determination, the number of TOAs included and the MJD
range covered, as well as the RMS of the post-fit timing
residuals. Asterisks indicate those pulsars which exhibit 
significant timing noise that has been removed, to first order,
by the fitting of a period second derivative.}
\begin{tabular}{lllccccl}
\hline
\multicolumn{1}{c}{PSR J} & \multicolumn{1}{c}{Period, $P$} &
\multicolumn{1}{c}{$\dot{P}$} &  
\multicolumn{1}{c}{Epoch} & \multicolumn{1}{c}{$N_{toa}$} & \multicolumn{1}{c}{Data Span} &
\multicolumn{1}{c}{Residual} & \multicolumn{1}{c}{DM}  \\
  & \multicolumn{1}{c}{(s)} & \multicolumn{1}{c}{($10^{-15}$)} &  
\multicolumn{1}{c}{(MJD)} &   & \multicolumn{1}{c}{(MJD)} &
\multicolumn{1}{c}{(ms)}  &  \multicolumn{1}{c}{(cm$^{-3}$ pc)} \\ 
\hline
0831$-$4406 & 0.311673518473(11)   & !!!1.279(6)    & 51396 &    38 & 51215--51577 &  0.47 & !254(5)           \\
0834$-$4159 & 0.121116349354(15)   & !!!4.437(8)    & 51505 &    27 & 51299--51710 &  1.88 & !240.5(15) \\    
0855$-$4644 & 0.064685817073(3)    & !!!7.2626(13)  & 51368 &    71 & 51158--52249 &  0.90 & !238.2(16) \\    
0855$-$4658 & 0.57507238258(4)     & !!13.625(20)   & 51395 &    39 & 51158--51577 &  1.17 & !472.7(12) \\    
0945$-$4833 & 0.3315855987755(18)  & !!!4.8291(6)   & 51703 &    35 & 51464--51940 &  0.15 & !!98.1(3) \\      
\\
\hline
\end{tabular}    
$^{\rm B}$ - binary pulsar, see Kaspi et al.~(2000b) for binary parameters.

\label{tb:prd}
\end{minipage}
\end{table*}

\clearpage
\addtocounter{table}{-1}
\begin{table*}
\begin{minipage}{150mm}
\caption{-- {\it continued}}
\begin{tabular}{lllccccl}
\hline
\multicolumn{1}{c}{PSR J} & \multicolumn{1}{c}{Period, $P$} &
\multicolumn{1}{c}{$\dot{P}$} &  
\multicolumn{1}{c}{Epoch} & \multicolumn{1}{c}{$N_{toa}$} & \multicolumn{1}{c}{Data Span} &
\multicolumn{1}{c}{Residual} & \multicolumn{1}{c}{DM}  \\
  & \multicolumn{1}{c}{(s)} & \multicolumn{1}{c}{($10^{-15}$)} &  
\multicolumn{1}{c}{(MJD)} &   & \multicolumn{1}{c}{(MJD)} &
\multicolumn{1}{c}{(ms)}  &  \multicolumn{1}{c}{(cm$^{-3}$ pc)} \\ 
\hline
1306$-$6242 & 0.98190211848(9)     & !!!5.86(3)     & 51448 &    25 & 51214--51680 &  2.21 & !480(6)  \\   
1309$-$6526 & 0.398292162785(20)   & !!!0.0184(15)  & 51767 &    32 & 51300--51958 &  1.67 & !340(4)  \\   
1314$-$6101 & 2.9483896038(15)     & !!11.7(9)      & 51396 &    22 & 51213--51577 &  7.05 & !309(13)  \\  
1319$-$6105 & 0.421118114249(6)    & !!!1.5012(11)  & 51709 &    27 & 51458--51958 &  0.28 & !442.2(5)  \\   
1322$-$6329 & 2.7642094208(5)      & !!11.08(10)    & 51456 &    22 & 51220--51691 &  2.80 & !659(9)  \\   
\\
\hline
\end{tabular}
\end{minipage}
\end{table*}

\clearpage
\addtocounter{table}{-1}
\begin{table*}
\begin{minipage}{150mm}
\caption{-- {\it continued}}
\begin{tabular}{lllccccl}
\hline
\multicolumn{1}{c}{PSR J} & \multicolumn{1}{c}{Period, $P$} &
\multicolumn{1}{c}{$\dot{P}$} &  
\multicolumn{1}{c}{Epoch} & \multicolumn{1}{c}{$N_{toa}$} & \multicolumn{1}{c}{Data Span} &
\multicolumn{1}{c}{Residual} & \multicolumn{1}{c}{DM}  \\
  & \multicolumn{1}{c}{(s)} & \multicolumn{1}{c}{($10^{-15}$)} &  
\multicolumn{1}{c}{(MJD)} &   & \multicolumn{1}{c}{(MJD)} &
\multicolumn{1}{c}{(ms)}  &  \multicolumn{1}{c}{(cm$^{-3}$ pc)} \\ 
\hline
1535$-$5848 & 0.307177662775(7)    & !!!2.716(3)    & 51760 &    20 & 51562--51957 &  !0.34! & !107(7)  \\   
1537$-$5153 &  1.52812410566(12)     & !!!4.172(12)    & 51423 &    31 & 51092--51753 &  !2.15! & !!93(4)  \\    
1538$-$5638 & 0.84398039688(14)    & !!!7.12(8)     & 51490 &    19 & 51299--51680 &  !2.53! & !546(7)  \\   
1538$-$5750 & 0.5065677(10)        & !!!0.042(3)    & 51032 &    34 & 51213--51753 &  !1.05! & !!91(4)  \\    
1539$-$5521 & 1.00495831813(8)     & !!!0.728(8)    & 51730 &    27 & 51300--51957 &  !1.66! & !380(5)  \\   
\\
\hline
\end{tabular}
\end{minipage}
\end{table*}

\clearpage
\addtocounter{table}{-1}
\begin{table*}
\begin{minipage}{150mm}
\caption{-- {\it continued}}
\begin{tabular}{lllccccl}
\hline
\multicolumn{1}{c}{PSR J} & \multicolumn{1}{c}{Period, $P$} &
\multicolumn{1}{c}{$\dot{P}$} &  
\multicolumn{1}{c}{Epoch} & \multicolumn{1}{c}{$N_{toa}$} & \multicolumn{1}{c}{Data Span} &
\multicolumn{1}{c}{Residual} & \multicolumn{1}{c}{DM}  \\
  & \multicolumn{1}{c}{(s)} & \multicolumn{1}{c}{($10^{-15}$)} &  
\multicolumn{1}{c}{(MJD)} &   & \multicolumn{1}{c}{(MJD)} &
\multicolumn{1}{c}{(ms)}  &  \multicolumn{1}{c}{(cm$^{-3}$ pc)} \\ 
\hline
1649$-$4653 & 0.55701875796(10)    & !!49.74(5)     & 51294 &    16 & 51089--51498 &  !2.27 & !332(4)  \\   
1650$-$4126 & 0.308917678787(3)    & !!!0.0198(4)   & 51685 &    26 & 51412--51956 &  !0.22 & !251.5(5)  \\   
1650$-$4341 & 0.309398365719(4)    & !!!0.01669(19) & 51737 &    20 & 51216--52257 &  !1.04 & !673(4)  \\   
1651$-$4519 & 0.51744320409(19)    & !!!8.19(11)    & 51399 &    20 & 51216--51581 &  !5.03 & !562(6)  \\   
1653$-$4315 & 0.41927974137(9)     & !!!0.015(4)    & 51587 &    38 & 50940--51554 &  !8.61 & !337(6)  \\   
\\
\hline
\end{tabular}
\end{minipage}
\end{table*}

\clearpage
\begin{table*}
\begin{minipage}{150mm}
\caption{\label{t:deriv} 
Derived parameters for 200 pulsars discovered
in the Parkes multibeam pulsar survey. We list the characteristic age,
the surface magnetic dipole field strength, the rate of rotational
energy loss, the distance derived from the DM and the TC93 model, the
inferred $z$-height and the corresponding radio luminosity at 1400
MHz.}
\begin{tabular}{lcccccc}
\hline
\multicolumn{1}{c}{PSR J} & log[$\tau_c$(yr)] & log[$B_s$(G)] & log[$\dot E$(erg s$^{-1}$] & 
Dist. & $z$ & Luminosity \\
 & & & & (kpc) & (kpc) & (mJy kpc$^2$) \\ 
\hline
0831$-$4406 &   6.59 &  11.81 &  33.22 & 12.7  &  $-$0.60 & !69.4 \\
0834$-$4159 &   5.64 &  11.87 &  34.99 & !9.7  &  $-$0.18 & !18.0 \\
0855$-$4644 &   5.15 &  11.84 &  36.03 & !9.9  &  $-$0.17 & !19.6 \\
0855$-$4658 &   5.83 &  12.45 &  33.45 & 28.3  &  $-$0.59 & 184.2 \\
0945$-$4833 &   6.04 &  12.11 &  33.72 & !2.7  &  ! 0.17 & !!2.9 \\
\\
\hline
\end{tabular}   
\label{tb:deriv}
\end{minipage}
\end{table*}

\clearpage
\addtocounter{table}{-1} 
\begin{table*}
\begin{minipage}{150mm}
\caption{-- {\it continued}}
\begin{tabular}{lcccccc}
\hline
\multicolumn{1}{c}{PSR J} & log[$\tau_c$(yr)] & log[$B_s$(G)] & log[$\dot E$(erg s$^{-1}$] & 
Dist. & $z$ & Luminosity \\
 & & & & (kpc) & (kpc) & (mJy kpc$^2$) \\ 
\hline
1306$-$6242 &   6.42 &  12.39 &  32.39 & 22.7  &   !0.05 & !72.1 \\
1309$-$6526 &   8.54 &  10.94 &  31.06 & 19.9  &  $-$0.91 & !59.4 \\
1314$-$6101 &   6.60 &  12.77 &  31.26 & !9.6  &   !0.29 & !37.7 \\
1319$-$6105 &   6.65 &  11.91 &  32.90 & 14.9  &   !0.41 & 186.5 \\
1322$-$6329 &   6.60 &  12.75 &  31.32 & 30.0  &  $-$0.43 & 153.0 \\
\\
\hline
\end{tabular}
\end{minipage}
\end{table*}

\clearpage
\addtocounter{table}{-1} 
\begin{table*}
\begin{minipage}{150mm}
\caption{-- {\it continued}}
\begin{tabular}{lcccccc}
\hline
\multicolumn{1}{c}{PSR J} & log[$\tau_c$(yr)] & log[$B_s$(G)] & log[$\dot E$(erg s$^{-1}$] & 
Dist. & $z$ & Luminosity \\
 & & & & (kpc) & (kpc) & (mJy kpc$^2$) \\ 
\hline
1535$-$5848 &   6.25 &  11.97 &  33.57 & !3.1  &  $-$0.13 & !!3.4 \\
1537$-$5153 &   6.76 &  12.41 &  31.67 & !2.6  &   !0.14 & !!0.2 \\
1538$-$5638 &   6.27 &  12.39 &  32.67 & 12.9  &  $-$0.20 & !46.6 \\
1538$-$5750 &   8.28 &  11.17 &  31.11 & !2.4  &  $-$0.08 & !!0.3 \\
1539$-$5521 &   7.34 &  11.94 &  31.45 & !6.8  &   !0.01 & !!6.6 \\
\\
\hline
\end{tabular}
\end{minipage}
\end{table*}

\clearpage
\addtocounter{table}{-1} 
\begin{table*}
\begin{minipage}{150mm}
\caption{-- {\it continued}}
\begin{tabular}{lcccccc}
\hline
\multicolumn{1}{c}{PSR J} & log[$\tau_c$(yr)] & log[$B_s$(G)] & log[$\dot E$(erg s$^{-1}$] & 
Dist. & $z$ & Luminosity \\
 & & & & (kpc) & (kpc) & (mJy kpc$^2$) \\ 
\hline
1649$-$4653 &   5.25 &  12.73 &  34.06 & !5.6  &  $-$0.14 & !!9.8 \\
1650$-$4126 &   8.39 &  10.90 &  31.42 & !5.1  &   !0.18 & !!7.4 \\
1650$-$4341 &   8.47 &  10.86 &  31.35 & !8.1  &   !0.07 & !17.1 \\
1651$-$4519 &   6.00 &  12.32 &  33.37 & !7.3  &  $-$0.09 & !28.7 \\
1653$-$4315 &   8.65 &  10.90 &  30.90 & !5.1  &   !0.03 & !13.9 \\
\\
\hline
\end{tabular}
\end{minipage}
\end{table*}

%%%%%%%%%%%%%%%%%%%%%%%%%%%%%%%%%%%%%%%%%%%%%%%%%%%%%
\clearpage
%%%%%%%%%%%%%%%%%%%%%%%%%%%%%%%%%%%%%%%%%%%%%%%%%%%%%%%%%%%%
\begin{table*}
\tabcolsep2pt
\caption{\label{t:egret} {\em EGRET} point sources as listed in the 3EG
catalogue (Hartman et al.~1999) and radio pulsars positionally
coincident with their error boxes. We list the {\em EGRET} source name
(column 1), and its identification as listed in the catalogue (column 2).
Assuming a spectral index
of 2, a flux value $\bar{F}$ ($E>100$ MeV)
in units of erg~s$^{-1}$~cm$^{-2}$ is
derived (column 3). A variability index, $V$, has been taken from
McLaughlin (2001) (column 4), and the size of the 95\% error-box is
quoted (column 5). Each {\em EGRET}
 point source is listed with pulsars (column 6)
where the relative positional difference (column 7) is less than or equal
to the size of the error-box. We also list those sources for which 
associations have been proposed in the literature.
 We quote the Taylor \& Cordes (1993) model
dispersion measure distances (column 8) and compare those to the new
NE2001 (Cordes \& Lazio 2002) distances (column 9), followed by the
characteristic age (column 10).  The NE2001 distance estimate is 
used to derive the $\dot{E}/d^2$ value (column 12) 
and the gamma-ray efficiency, $\eta$, for a beaming fraction of 
$1/4\pi$ (column 13), using the quoted
spin-down luminosity (column 11). The likelihood of a genuine
association is indicated by a `+', `$-$', `+?' or `?' in column
14. References to the pulsar and/or already proposed associations are
given in column 15.}
%\begin{tabular}{ccr@{$\pm$}lcccccrrcc@{\hspace{-2em}}rcl}
\begin{tabular}{ccr@{$\pm$}lcccccrrccrcl}
\hline
\hline
\noalign{\medskip}
3EG J & ID 
      & \multicolumn{2}{c}{$\bar{F}$} & $V$ & $\theta_{95}$
      & PSR & $\Delta\Phi/\theta_{95}$ & \multicolumn{1}{c}{$d_{\rm TC93}$}
      & \multicolumn{1}{c}{$d_{\rm ne01}$}
      & \multicolumn{1}{c}{$\tau$}
      & \multicolumn{1}{c}{$\log[\dot{E}$ (erg s$^{-1}$)]}
      & \multicolumn{1}{c}{$\log[\dot{E}$ (erg
                                         }
      & $\eta_{f=1/4\pi}$
      & Q & Ref \\
      & \multicolumn{4}{c}{\quad($10^{-10}$erg s$^{-1}$cm$^{-2}$)}\hspace{1em}
       & (deg) &  & 
      & \multicolumn{1}{c}{(kpc)} 
      & \multicolumn{1}{c}{(kpc)} 
      & \multicolumn{1}{c}{(kyr)} 
      & \multicolumn{1}{c}{ } 
      & \multicolumn{1}{c}{
                         s$^{-1}$kpc$^{-2}$)]}
      & (\%)!
      & & \\ 
\noalign{\smallskip}
\hline
\noalign{\smallskip}
0222+4253 & P &  1.39 & 0.22 & 0.13 & 0.31 & J0218+4232 & 0.93 
  & 5.9 & 2.9 & 490000 & 35.38 & 34.53 & 3.9! & + & 1 \\ 
\noalign{\smallskip}
0500$-$0159 & A & 0.83 & 0.17 & 3.63 & 0.75 & J0459$-$0210 & 0.30
  & 1.3 & 0.9 & 13000 &   31.60 & 31.70 & 1800!!! & $-$ & 2 \\
\noalign{\smallskip}
0533$-$6916 & G & 1.06 & 0.16 & 1.41 & 0.53 & B0540$-$69!! & 1.09 
  & 49.4$^\ast$ & $>47.8$ & 1.8 & 38.12 & 34.82 & 1.5! & $-$ & 3 \\
       &   & \multicolumn{2}{c}{ } & & & J0535$-$6935 & 0.64
  & 49.4$^\ast$ & $>47.3$ & 397 & 34.60 & 31.30 & 28.6! & $-$ & 4  \\ 
\noalign{\smallskip}
0534+2200 & P &  16.84 & 0.35 & 0.50 & 0.05 & B0531+21!! & 1.23 
 & 2.5$^\ast$ & 1.7 & 1.3 & 38.64 & 38.04 & 0.02 & + & 5 \\
\noalign{\smallskip}
0834$-$4511 & P & 62.10 & 0.83 & 0.96 & 0.02 & B0833$-$45!! 
 & 3.72 & 0.5$^\ast$ & 0.2 & !11 & 36.84 & 37.44 & 0.2! &  + & 6 \\
\noalign{\smallskip}
1013$-$5915 & ? & 2.49 & 0.45 & 0.15 & 0.72 & B1011$-$58!! 
 & 0.41 & 10.2 & 7.9 & !21 & 33.11 & 31.30 & 11500!!! & $-$ & 7 \\
  & & \multicolumn{2}{c}{ } & & & J1013$-$5934
  & 0.45 & 11.3 & 8.5 & 12580 & 32.48 & 30.48 & 69000!!! & $-$ & III  \\
  & & \multicolumn{2}{c}{ } & & & J1016$-$5857 
  & 0.67 & 9.3 & 8.0 & !21 & 36.42 & 34.61 & 5.8! & +? & I, 8 \\
\noalign{\smallskip}
1014$-$5705 & ? & 2.53 & 0.48 & 0.55 & 0.67 & J1015$-$5719
  & 0.44 & 4.9 & 5.1 & !39 & 35.92 & 34.50 & 7.5! & +? & III, 9  \\
\noalign{\smallskip}
1048$-$5840 & ? & 4.60 & 0.50 & 0.01 & 0.17 & B1046$-$58!!
  & 0.90 & 3.0 & 2.7 & !20 & 36.30 & 35.44 & 1.6! & + & 10, 20 \\
\noalign{\smallskip}
1058$-$5234 & P & 2.48 & 0.28 & 0.94 & 0.25 & B1055$-$52!! 
 & 0.66 & 1.5 & 0.7 & 535 & 34.48 & 34.79 & 4.1! & + & 11 \\
\noalign{\smallskip}
1102$-$6103 & ? & 2.42 & 0.46 & 2.38 & 0.61 & J1104$-$6103
  & 0.37 & 2.3 & 1.9 & 2263 & 33.54 & 33.00 & 250!!! & $-$ & 12 \\
  & &   \multicolumn{2}{c}{ } & & & J1105$-$6107
  & 0.62 & 7.1 & 5.0 & !63 & 36.40 & 35.00 & 2.3! & + & 10, 13 \\
\noalign{\smallskip}
1308$-$6112 & ? & 1.64 & 0.45 & 4.61 & 0.71 & J1314$-$6101
  & 1.00 & 9.6 & 6.1 & 3891 & 31.30 & 29.70 & 3000000!!! & $-$ & III \\
\noalign{\smallskip}
1410$-$6147 & ? &  4.78 & 0.65 & 0.72 & 0.36 & J1412$-$6145
  & 0.41 & 9.3 & 7.8 & !51 & 35.08 & 33.30 & 230!!! & $-$ & I, 14 \\
  & &  \multicolumn{2}{c}{ } & & & J1413$-$6141
  & 0.75 & 11.0 & 10.1 & !14 & 35.75 & 33.74 & 84!!! & $?$ & III, 9, 14 \\
\noalign{\smallskip}
1420$-$6038 & ? & 3.33 & 0.64 & 1.14 & 0.32 &  J1420$-$6048 
 & 0.53 & 7.7 & 5.6 & !13 & 37.00 & 35.50 & 1.0! & +? & III, 15\\
\noalign{\smallskip}
1638$-$5155 & ? & 2.23 & 0.45 & 1.23 & 0.68 & J1638$-$5226
 & 0.79 & 4.9 & 3.3 & 2035 & 33.43 & 32.48 & 970!!! & $-$ & III \\
\noalign{\smallskip}
1639$-$4702 & ? & 3.96 & 0.65 & 1.88 & 0.56 & B1636$-$47!! 
 & 0.52 & 7.3 & 6.5 & 195 & 34.08 & 32.48 & 1300!!! & $-$ & 7 \\
  & &  \multicolumn{2}{c}{ } & & & J1637$-$4642
 & 0.83 & 5.8 & 5.1 & !41 & 35.81 & 34.39 & 15.2! & ? & III, 9 \\
  & &  \multicolumn{2}{c}{ } & & & J1637$-$4721
 & 0.81 & 5.9 & 5.3 & 4156 & 32.00 & 30.60 & 94000!!! & $-$ & III \\
  & &  \multicolumn{2}{c}{ } & & & !J1638$-$4715$^\dagger$
 & 0.46 & 6.7 & 6.1 & 59 & 33.56 & 32.00 & 3800!!!! & $-$ & IV \\
 & &   \multicolumn{2}{c}{ } & & & J1640$-$4648
 & 0.65 & 6.1 & 5.5 & 3506 & 33.75 & 32.30 & 2030!!! & $-$ & III \\
\noalign{\smallskip}
1704$-$4732 & ? & 1.53 & 1.53 & 7.73 & 0.66 & J1707$-$4729 
 & 0.71 & 10.5 & 6.2 & 2706 & 33.52 & 31.95 & 1660!!! & $-$ & III \\
\noalign{\smallskip}
1710$-$4439 & P & 8.35 & 0.46 & 0.03 & 0.09 & B1706$-$44!!
 & 2.26 & 1.8 & 2.3 & !18 & 36.52 &  35.81 & 1.2! & $+$ & 16 \\
\noalign{\smallskip}
1714$-$3857 & ? & 3.25 & 0.48 & 2.27 & 0.51 & J1713$-$3844
 & 0.59 & 6.5 & 5.9 & 143 & 33.23 & 31.70 & 7630!! & $-$ & III, 9 \\
  & &   \multicolumn{2}{c}{ } & & & J1715$-$3903
 & 0.48 & 4.8 & 4.1 & 117 & 34.84 & 33.61 & 74.3! & $-$ & III, 9 \\
\noalign{\smallskip}
1736$-$2908 & ? &  2.46 & 0.44 & 7.73 & 0.62 & J1736$-$2843
 & 0.67 & 7.4 & 5.8 & 3400 & 30.60 & 29.00 & 18000!!! & $-$ & IV \\
\noalign{\smallskip}
1741$-$2050 & ? &  1.79 & 0.29 & 4.24 & 0.63 & J1741$-$2019
 & 0.84 & 2.0 & 1.72 & 3805 & 31.00 & 30.48 & 46000!!! & $-$ & 17 \\
\noalign{\smallskip}
1746$-$1001 & ? &  1.47 & 0.26 & 9.30 & 0.76 & J1745$-$0952
 & 0.84 & 2.4 & 1.8 & 3200000 & 32.70 & 32.20 & 900!!! & $-$ & 18 \\
\noalign{\smallskip}
1824$-$1514 & ? &  2.62 & 0.48 & 3.59 & 0.52 & B1822$-$14!!
 & 0.89 & 5.4 & 5.0 & 195 & 34.61 & 33.21 & 154!!! & $-$ &  19 \\
 & &  \multicolumn{2}{c}{ } & & & J1823$-$1526
 & 0.77 & 9.3 & 8.1 & 6092 & 31.60 & 29.78 & 420000!!! & $-$ & II \\
 & &  \multicolumn{2}{c}{ } & & & J1824$-$1505
 & 0.38 & 8.6 & 7.5 & 4687 & 32.90 & 31.00 & 18000!!! & $-$ & IV \\
 & &  \multicolumn{2}{c}{ } & & & J1826$-$1526
 & 0.75 & 10.9 & 8.2 & 5581 & 32.90 & 31.00 & 22000!!! & $-$ & II \\
\noalign{\smallskip}
\hline
\end{tabular}
\end{table*}

\addtocounter{table}{-1}
\begin{table*}
\tabcolsep2pt
\caption{$-$ {\it continued}}
\begin{tabular}{ccr@{$\pm$}lcccccrrccrcl}
\hline
\hline
\noalign{\medskip}
\noalign{\medskip}
3EG J & ID 
      & \multicolumn{2}{c}{$\bar{F}$} & $V$ & $\theta_{95}$
      & PSR & $\Delta\Phi/\theta_{95}$ & \multicolumn{1}{c}{$d_{\rm TC93}$}
      & \multicolumn{1}{c}{$d_{\em ne01}$}
      & \multicolumn{1}{c}{$\tau$}
      & \multicolumn{1}{c}{$\log[\dot{E}$ (erg s$^{-1}$)]} 
      & \multicolumn{1}{c}{$\log[\dot{E}$ (erg
                                         }
      & $\eta_{f=1/4\pi}$
      & Q & Ref \\
      & \multicolumn{4}{c}{\quad($10^{-10}$erg s$^{-1}$cm$^{-2}$)}\hspace{1em}
       & (deg) &  & 
      & \multicolumn{1}{c}{(kpc)} 
      & \multicolumn{1}{c}{(kpc)} 
      & \multicolumn{1}{c}{(kyr)} 
      & \multicolumn{1}{c}{ } 
      & \multicolumn{1}{c}{
                          s$^{-1}$kpc$^{-2}$)]}
      & (\%)!
      & & \\ 
\noalign{\smallskip}
\hline
\noalign{\smallskip}
1826$-$1302 & ? &  3.45 & 0.54 & 5.89 & 0.46 & B1823$-$13!!
 & 1.17 & 4.1 & 3.9 & !21 & 36.46 & 35.28 & 1.8 & $?$ & 19, 20 \\
\noalign{\smallskip}
1837$-$0423 & ? & 1.42 & 1.42 & 9.85 & 0.52 & B1834$-$04!!
 & 0.55 & 4.6 & 4.9 & 3381 & 33.18 & 31.78 & 2200!! & $-$ & 19 \\
 &  &   \multicolumn{2}{c}{ } & & & J1838$-$0453
 & 0.98 & 8.3 & 8.1 & !52 & 34.92 & 33.11 & 110!! & $-$ & II \\
\noalign{\smallskip}
1837$-$0606 & ? & 3.69 & 0.59 & 2.81 & 0.19 & J1837$-$0559
 & 0.74 & 5.0 & 5.4 & 964 & 34.20 & 32.78 & 650!! & $-$ & II \\
 & &   \multicolumn{2}{c}{ } & & & J1837$-$0604
 & 0.90 & 6.2 & 6.4 & !34 & 36.30 & 34.69 & 7.2 & +? & II, 15 \\
\noalign{\smallskip}
1850$-$2652 & ? & 0.48 & 0.19 & 4.50 & 1.00 & J1852$-$2610
 & 0.88 & 2.2 & 1.8 & 610000 & 32.00 & 31.48 & 1600!! & $-$ & 2 \\
\noalign{\smallskip}
1856+0114 & ? & 5.02 & 0.64 & 1.32 & 0.19 & B1853+01!!
 & 0.30 & 2.8 & 3.1 & !20 & 35.63 & 34.65 & 10.5 & ? &  19, 21 \\
\noalign{\smallskip}
1903+0550 & ? & 4.62 & 0.66 & 3.14 & 0.64 & B1900+05!!
 & 0.39 & 3.9 & 4.7 & 917 & 33.08 & 31.70 & 8100!! & $-$ & 22  \\
 & &   \multicolumn{2}{c}{ } & & & J1903+0609
 & 0.31 & 8.1 & 7.1 & 308 & 34.15 & 32.48 & 1600!! & $-$ & IV \\
 & &   \multicolumn{2}{c}{ } & & & J1905+0603
 & 0.75 & 18.3 & 11.7 & 7807 & 32.60 & 30.48 & 150000!! & $-$ & IV \\
 & &  \multicolumn{2}{c}{ } & & & J1905+0616
 & 0.88 & 5.3 & 5.7 & 116 & 33.74 & 32.30 & 2600!! & $-$ & II  \\
\noalign{\smallskip}
2021+3716 & ? & 4.40 & 0.46 & 2.53 & 0.30 & J2021+3651
 & 1.38 & 19.2 & 12.4 & 17 & 36.53 & 34.34 & 18.9 & +? & 23 \\
\noalign{\smallskip}
2227+6122 & ? & 41.3 & 6.1 & 0.19 & 0.49 & J2229+6114
 & 0.56 & 12.0 & 7.3 & 11 & 37.34 & 35.62 & 0.7 & $+$ & 24 \\
\noalign{\smallskip}
\hline
\end{tabular}

{\flushleft
$^\ast$ Distance obtained by independent estimate, used for $\dot{E}/d^2$ \\
$^\dagger$ Name and parameters based on prelimanary timing solution.\\
References. 
III -- this work,
I -- Paper I, II -- Paper II,
IV -- Parkes Pulsar, unpublished \\
1 -- Kuiper et al.~(2000), 2 -- D'Amico et al.~(1998), 
3 -- Manchester et al.~(1993),
4 -- Crawford et al.~(2001), 5 -- Staelin \& Reifenstein (1968), 
6 -- Large et al.~(1968), 7 -- Johnston et al.~(1992), 
8 -- Camilo et al.~(2001a),
9 -- Torres et al.~(2001),
10 -- Kaspi et al.~(2000), 
11 -- Fierro et al.~(1993), 
12 -- Kaspi et al.~(1996), 
13 -- Kaspi et al.~(1997), 
14 -- Manchester et al.~(2002), 
15 -- D'Amico et al.~(2001), 
16 -- Thompson et al.~(1992), 
17 -- Edwards et al.~(2001),  
18 -- Edwards \& Bailes (2001), 
19 -- Clifton et al.~(1992), 
20 -- Merck et al.~(1996), 
21 -- Wolszczan et al.~(1991),
22 -- Hulse \& Taylor (1974), 
23 -- Roberts et al.~(2002)
24 -- Halpern et al.~(2001)
}

\end{table*}
%%% References in table %%%
%\nocite{hbb+99}
\nocite{khv+00}
\nocite{dsb+98}
\nocite{mml+93}
\nocite{ckm+01}
\nocite{sr68}
% 5
\nocite{lvm68}
\nocite{jml+92}
\nocite{cbm+01}
\nocite{klm+00}
\nocite{fbb+93}
% 10
\nocite{kbm+97}
\nocite{mbc+02}
\nocite{tbc01}
\nocite{dkm+01}
\nocite{clj+92}
% 15
\nocite{mlt+78}
\nocite{ht74}
\nocite{kmj+96}
\nocite{tab+92}
\nocite{ebvb01}
\nocite{eb01b}
\nocite{mbd+96}
\nocite{wcd91}
\nocite{rhr+02}
\nocite{hcg+01}

%%%%%%%%%%%%%%%%%%%%%%%%%%%%%%%%%%%%%%%%%%%%%%%%%%%%%%%

\clearpage
%%%%%%%%%%%%%%%%%%%%%%%%%%%%%%%%%%%%%%%%%%%%%%%%%%%%%%%
\begin{table}
\caption{Young pulsars with spin parameters similar to the Vela pulsar. All sources
selected have ages $10\le \tau_c \le 100$ kyr
and spin-down luminosities of 
$\dot{E}\ge 10^{36}$ erg s$^{-1}$.
Pulsars marked with G are known to exhibit glitches, while
sources labelled with E are or may be associated with {\em EGRET\/}
point sources. \label{t:velalike}}

\begin{tabular}{l@{\hspace{1em}}rrcrc@{\hspace{-0.5em}}l@{\hspace{0.2em}}l}
\hline
\hline
\noalign{\medskip}
\multicolumn{1}{c}{PSR} & \multicolumn{1}{c}{$P$} 
  & \multicolumn{1}{c}{Age, $\tau_c$}
  & \multicolumn{3}{c}{$\dot{E}$} & Ref & Notes \\
  & \multicolumn{1}{c}{(ms)} 
  & \multicolumn{1}{c}{(kyr)}
  & \multicolumn{3}{c}{($10^{36}$ erg s$^{-1}$)} &  &  \\
\noalign{\smallskip}
\hline
\noalign{\smallskip}
%%  PSR           Period     Age           Edot
 B0833$-$45  &    89.3  &  11 & \hspace{1em}
 &  6.9 & &  1 & G, E  \\
 J0855$-$4644  &  64.7  & 141 & &  1.1& & III & \\
 J0940$-$5428  &  87.5  & 42 &  &  1.9& & I & \\
 J1016$-$5857  &  107.4 &  21  & &  2.6& & 2 & G, E\\
 B1046$-$58    &  123.7  &  20  & & 2.0& &  3, 4  & G, E\\
\noalign{\smallskip}
 J1105$-$6107  &  63.2   &  63  & & 2.5& & 5 & G, E \\
 J1112$-$6103  &  65.0   &  33  & & 4.5& & I & \\
% J1119$-$6127  &  407.6 & 1.6 & & 2.3 & & 6 & G \\
 J1301$-$6305  &  184.5 &  11  & & 1.7& & I & \\
 B1338$-$62    &  193.3 &  12  & & 1.4& & 7 & G\\
 J1420$-$6048  &  68.2  &  13  & & 10.0&& III, 8 & E \\
\noalign{\smallskip}
 J1524$-$5625  &  78.2  &  32  & & 3.2& & III & \\
 J1531$-$5610  &  84.2  &  97  & & 0.9& & III & \\
 B1706$-$44    &  102.5 &  18  & & 3.4& & 3 & G, E\\
 J1718$-$3825  &  74.7  &  90  & & 1.3& & I & \\
 B1727$-$33    &  139.4 &  26  & & 1.2& & 3 & G\\
\noalign{\smallskip}
 J1747$-$2958  &  98.8  &  26  & & 2.5& & 9 &  \\
 B1757$-$24    &  124.9 &  16  & & 2.6& & 7 & G \\
 B1800$-$21    &  133.6 &  16  & & 2.2& & 10 & G \\
 J1809$-$1917  &  82.7  &  51  & & 1.8& & II &  \\
% J1811$-$1926  &  64.7 &  23  & & 6.4& & 65-ms X-ray psr \\
 B1823$-$13    &  101.5 &  21  & & 2.9& & 10 & G, E \\
\noalign{\smallskip}
 J1828$-$1101  &  72.1  &  77  & & 1.6& & II & \\
 J1837$-$0604  &  96.3  &  34  & & 2.0& & II, 8 &\\
 J1913+1011    &  35.9  &  169 & & 2.9& & I & \\
 B1951+32      &  39.5 &  107 & & 3.7& & 11 & \\
 J2021+3651    & 103.7  &  17 &  &  3.4 & & 12 & E \\
\noalign{\smallskip}
 J2229+6114    &  51.6 &  11 & & 22.0 & & 13 & E \\
\hline
\end{tabular}
{\flushleft
References. 
III -- this work,
I -- Paper I,
II -- Paper II,
1 -- Large et al.~(1968),
2 -- Camilo et al.~(2001a),
3 -- Johnston et al.~(1992),
4 -- Kaspi et al.~(2000a),
5 -- Kaspi et al.~(1997),
6 -- Camilo et al.~(2000),
7 -- Manchester et al.~(1985),
8 -- D'Amico et al.~(2001), 
9 -- Camilo et al.~(2002a),
10 -- Clifton et al.~(1992),
11 -- Kulkarni et al.~(1988)
12 -- Roberts et al.~(2002)
13 -- Halpern et al.~(2001)
}

\end{table}
\nocite{lvm68}
\nocite{jml+92}
\nocite{klm+00}
\nocite{kbm+97}
\nocite{ckl+00}
\nocite{mdt85}
\nocite{dkm+01}
\nocite{cmgl02}
\nocite{clj+92}
\nocite{kcb+88}
%%%%%%%%%%%%%%%%%%%%%%%%%%%%%%%%%%%%%%%%%%%%%%%%%%%%%%%

%%%%%%%%%%%%%%%%%%%%%%%%%%%%%%%%%%%%%%%%%%%%%%%%%%%

\clearpage
\begin{figure*} 
\centerline{\psfig{file=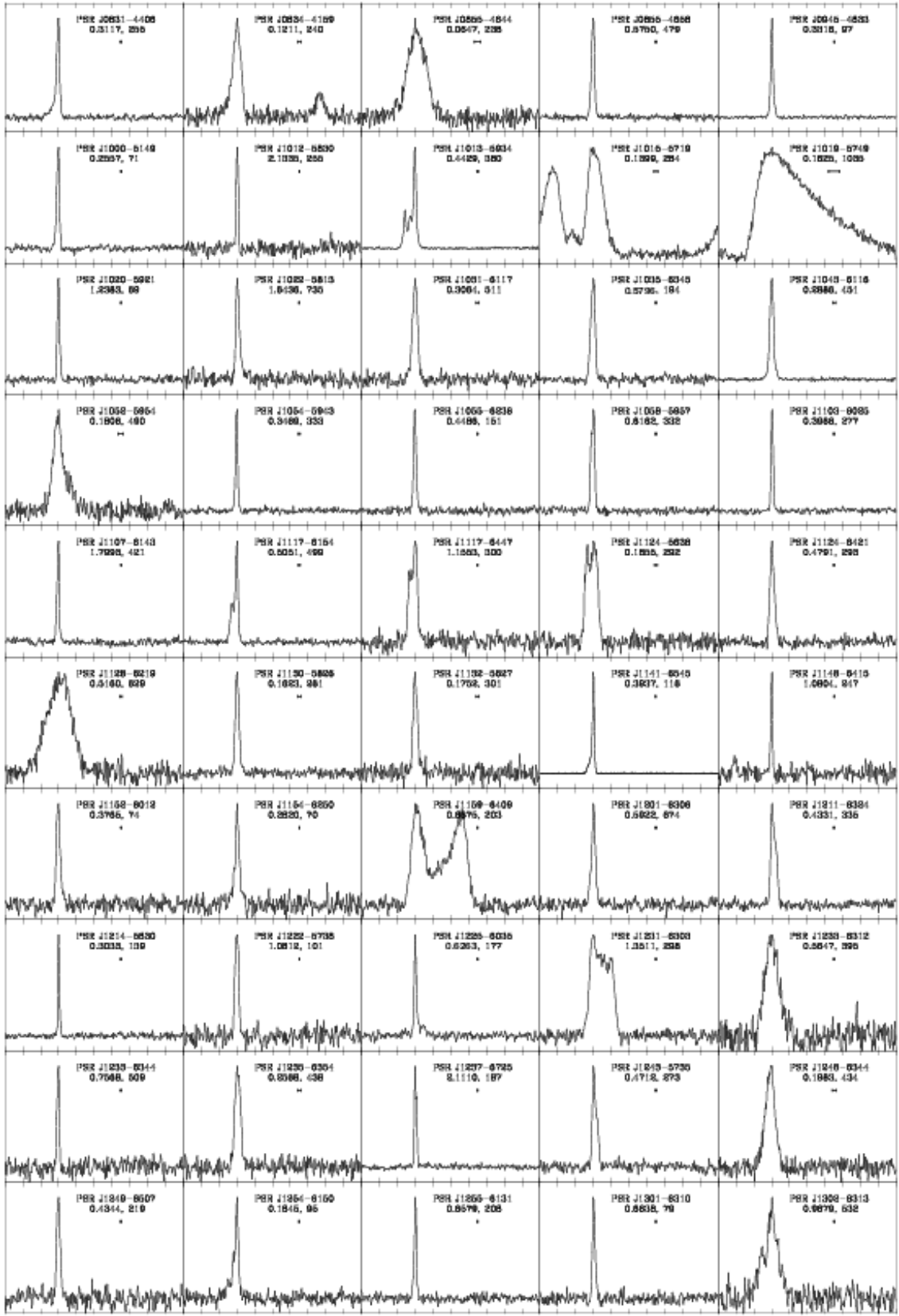,height= 220mm}} 
\caption{Mean pulse profiles at 1374MHz for 200 pulsars discovered in the
Parkes multibeam survey. The highest point in the profile is placed at
phase 0.3. For each profile, the pulsar Jname, pulse period 
and dispersion measure are
given. The small horizontal bar under the period indicates the
effective resolution of the profile, including the effects of
interstellar dispersion.}
\label{fg:prf}
\end{figure*}

\clearpage
\addtocounter{figure}{-1}
\begin{figure*} 
\centerline{\psfig{file=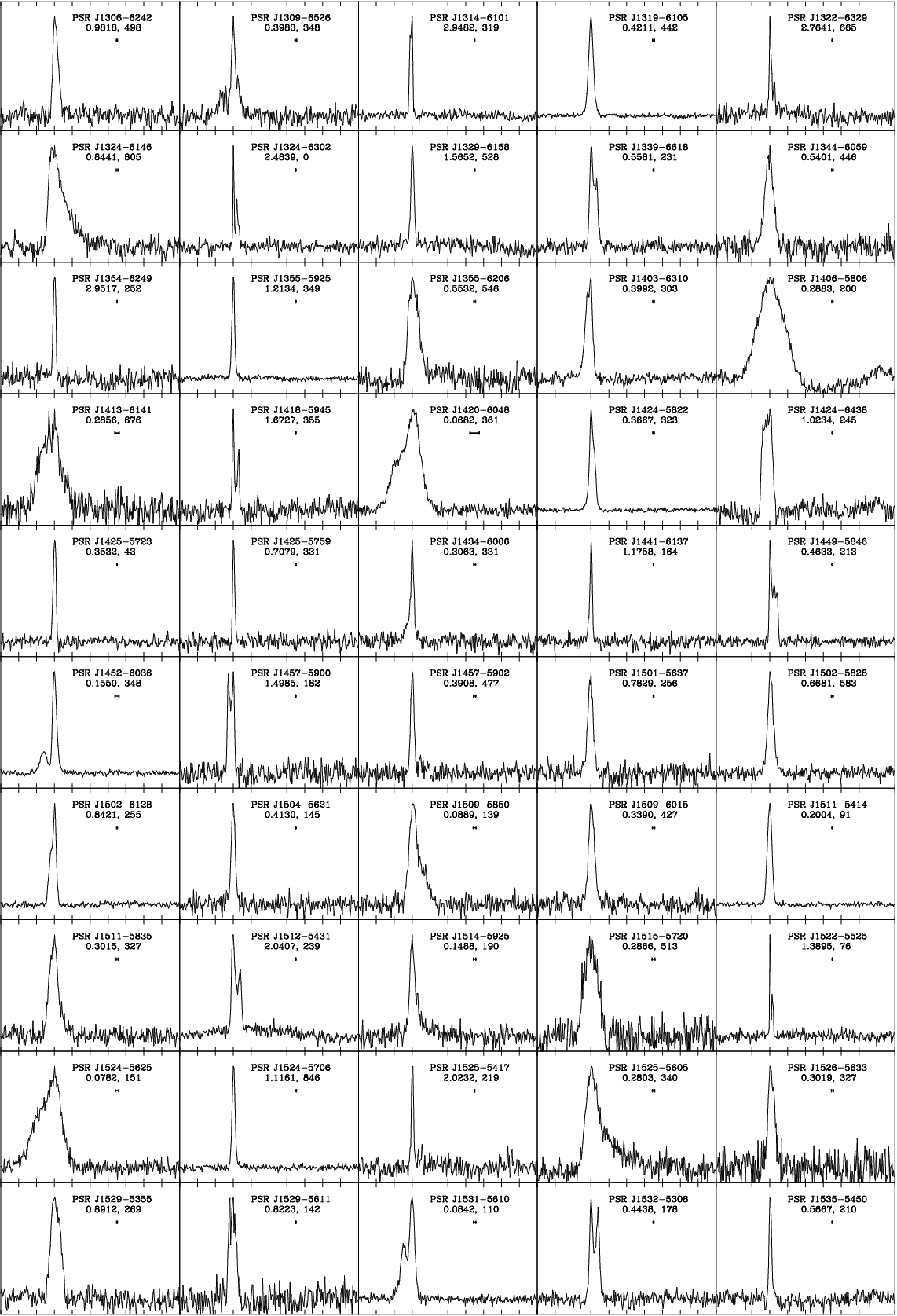,height=220mm}} 
\caption{-- {\it continued}}
\end{figure*}

\clearpage
\addtocounter{figure}{-1}
\begin{figure*} 
\centerline{\psfig{file=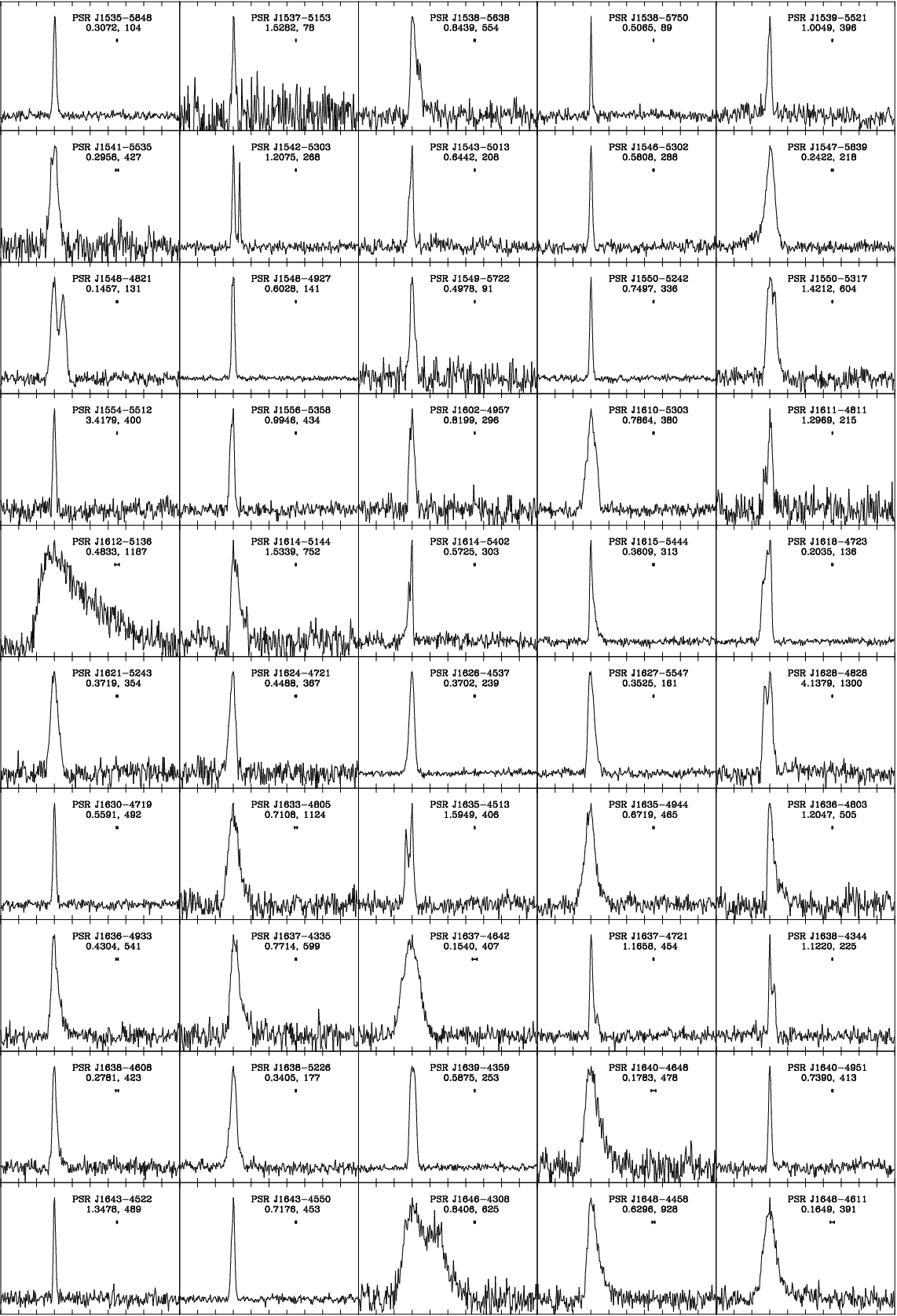,height=220mm}} 
\caption{-- {\it continued}}
\end{figure*}

\clearpage
\addtocounter{figure}{-1}
\begin{figure*} 
\centerline{\psfig{file=prfs4.ps,height=220mm}} 
\caption{-- {\it continued}}
\end{figure*}
\clearpage
%%%%%%%%%%%%%%%%%%%%%%%%%%%%%%%%%%%%%%%%%%%%%%%%%%%%%%%%%%

\begin{figure}

\centerline{\psfig{file=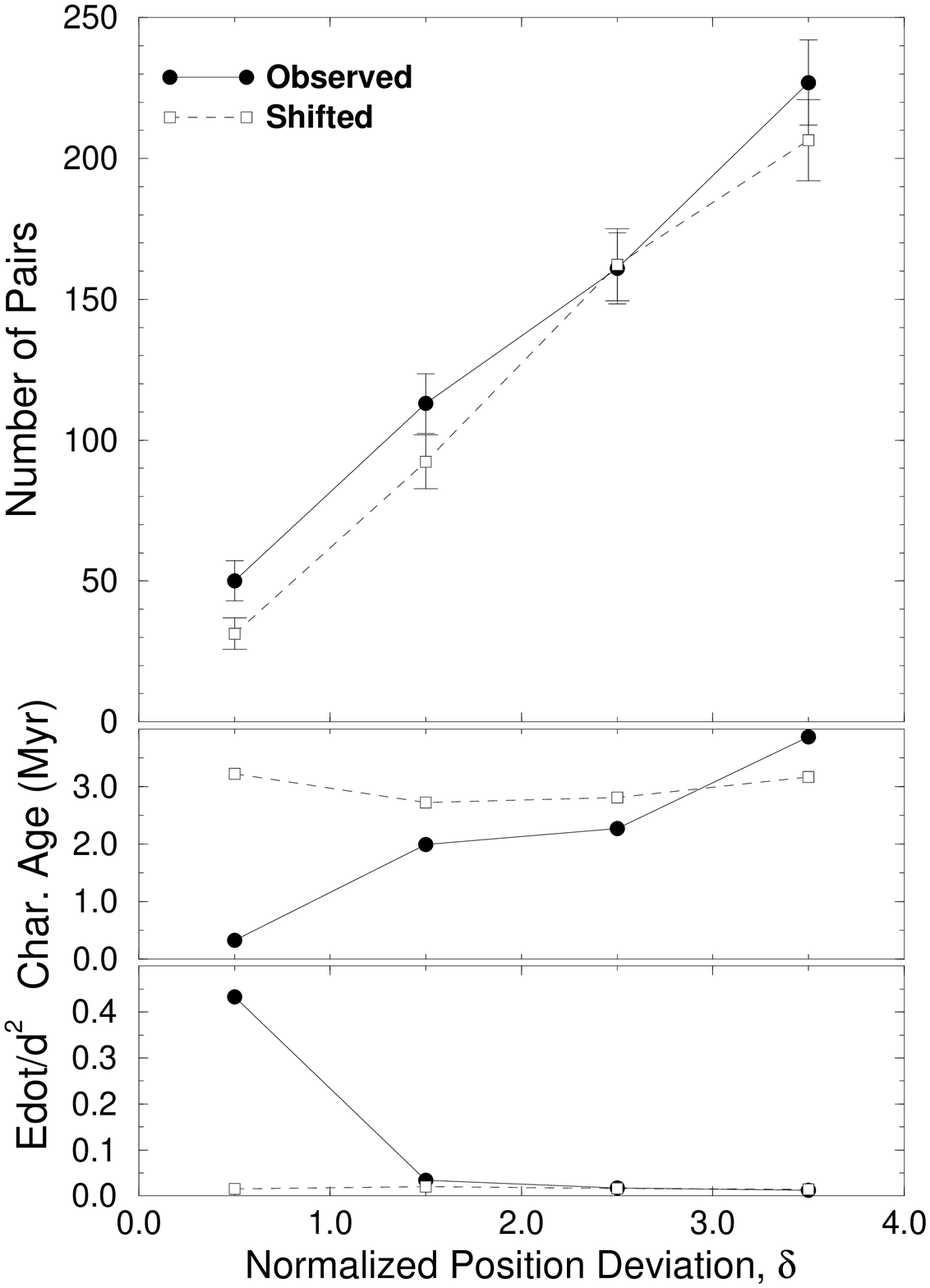,width=9cm}}

\caption{\label{f:offsetplot}
{\em Top panel:} Number of {\em EGRET\/} source/pulsar pairs as 
found for the observed sample of pulsars, and a sample
of pulsars shifted systematic in Galactic longitude,
as a function of normalized position deviation. Points
are plotted in centre of the annulus intervals $\delta + d\delta$
considered.
{\em Middle panel:} Median characteristic age of the pulsars
in the corresponding pairs, {\em Bottom panel:} median $\dot{E}/d^2$
of the pulsars in the corresponding pairs.
}
\end{figure}

\begin{figure}

\centerline{\psfig{file=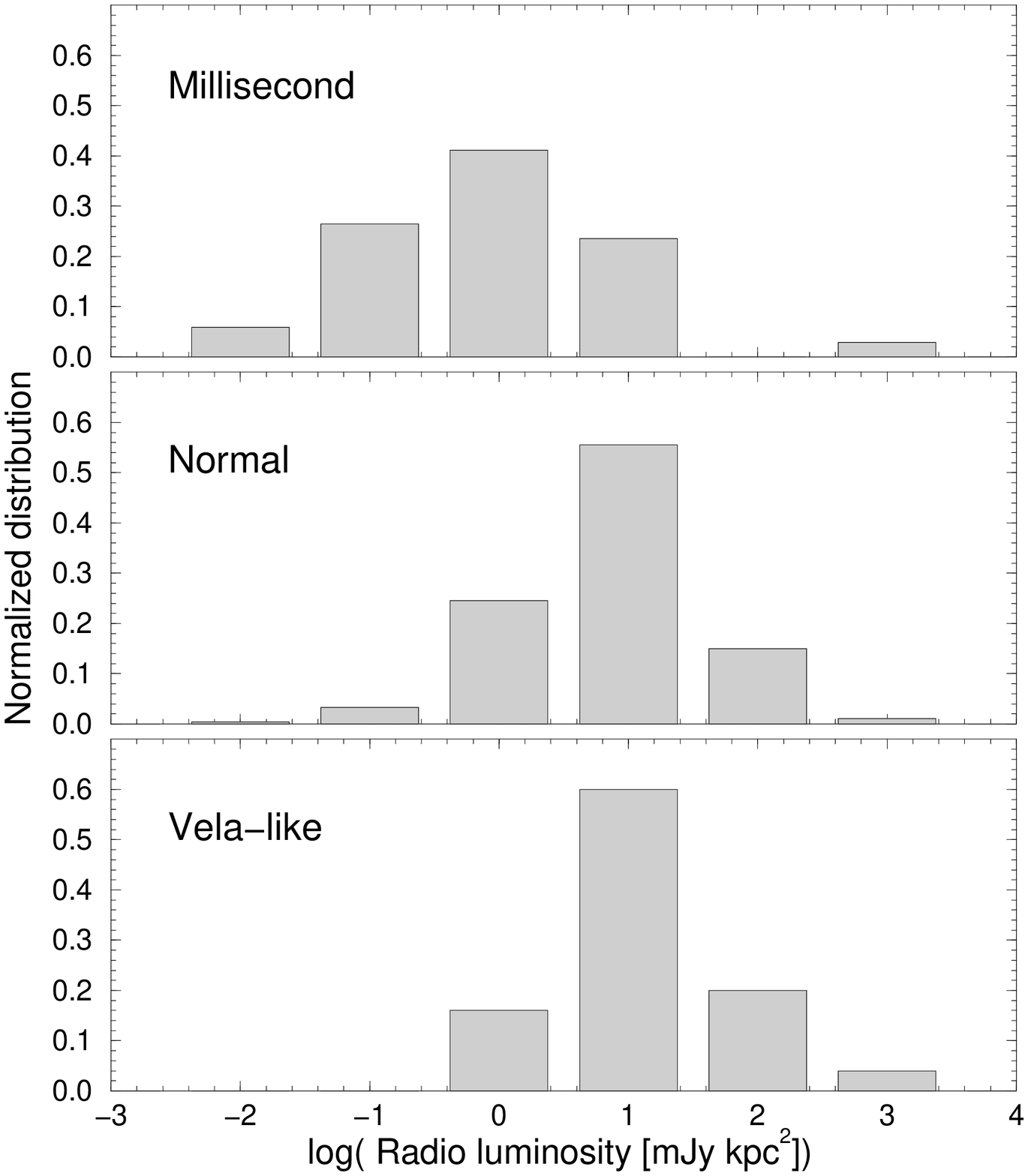,width=9cm}}

\caption{\label{f:radiolum}
Observed distribution of radio luminosities measured
at 1400 MHz for millisecond pulsars, `normal' pulsars and
`Vela-like' pulsars.}
\end{figure}

\begin{figure}

\centerline{\psfig{file=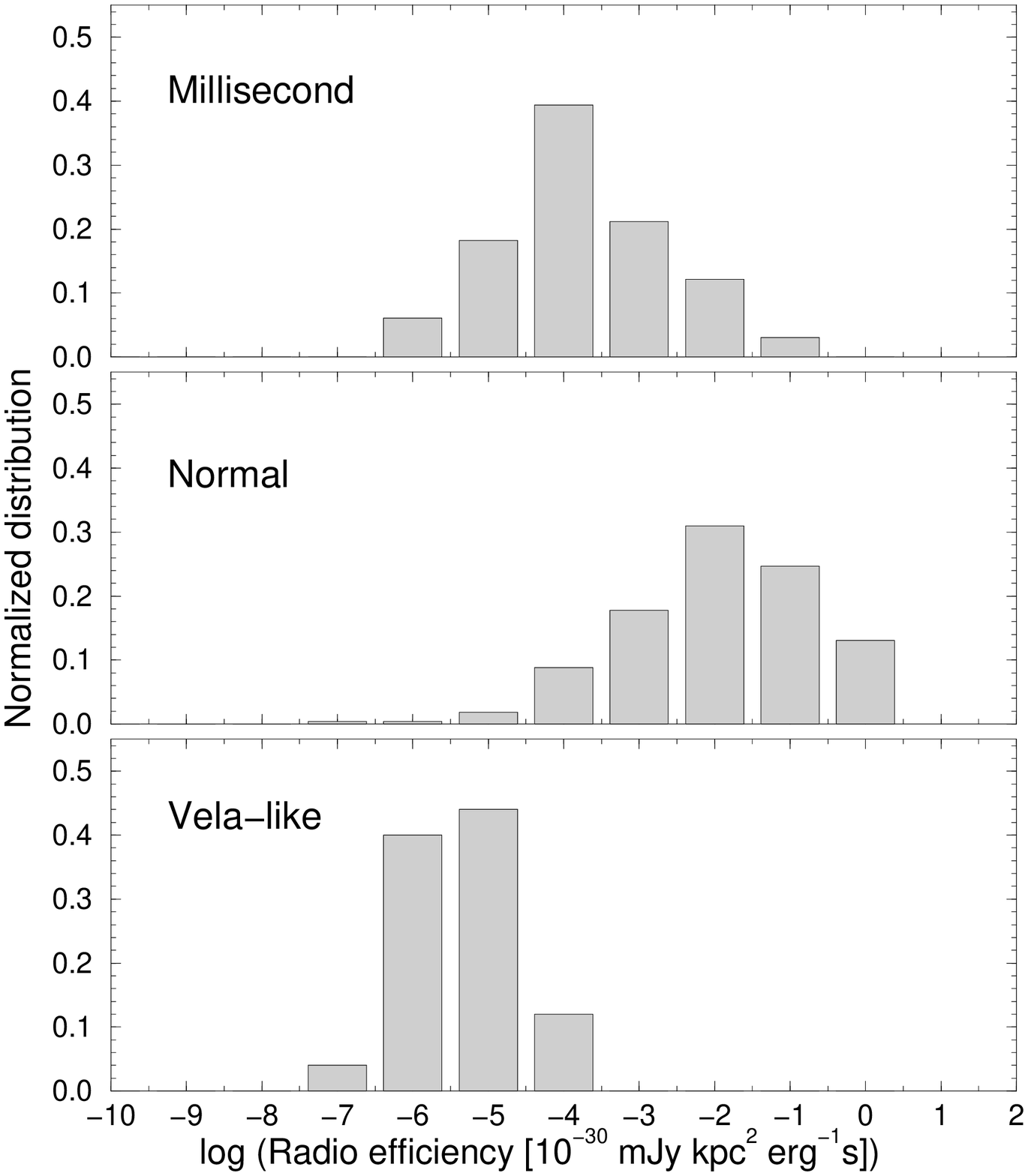,width=9cm}}

\caption{\label{f:radioeff}
Observed distribution of radio efficiencies measured
at 1400 MHz for millisecond pulsars, normal pulsars and
Vela-like pulsars.}
\end{figure}

\begin{figure}

\centerline{\psfig{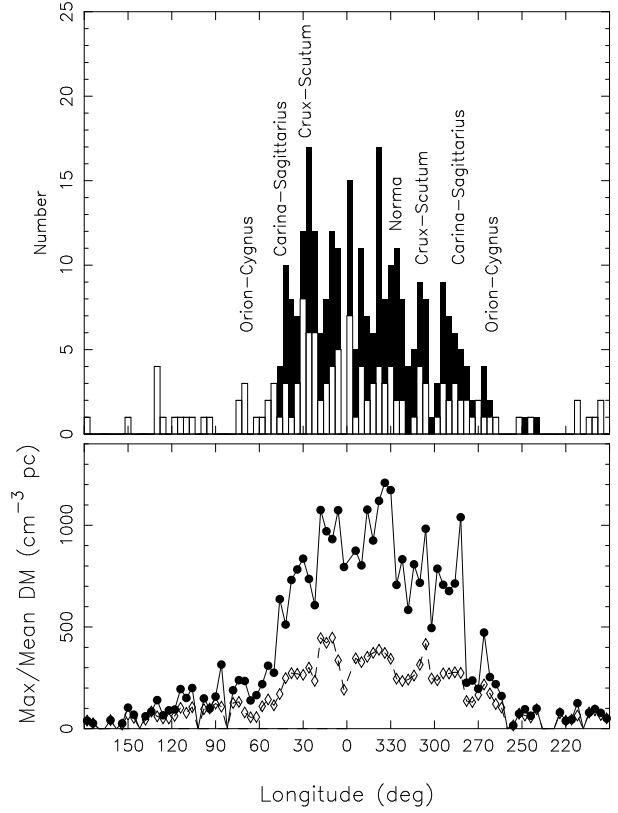}}

\caption{\label{f:lnum}\label{f:maxmeandm} 
{\em Top:} Number of pulsars with Galactic latitude $|b|\le20^\circ$ and
a characteristic age of less then 1 Myr as a function of
Galactic longitude. Filled bars mark the numbers of
pulsars from the PMPS. The locations of
the lines-of-sight tangential to the Galactic spiral
arms are indicated. 
{\em Bottom:}
Maximum (filled circles) and mean (open diamonds) dispersion
measures for these pulsars as a function of Galactic longitude.}
\end{figure}

\begin{figure}

\centerline{\psfig{file=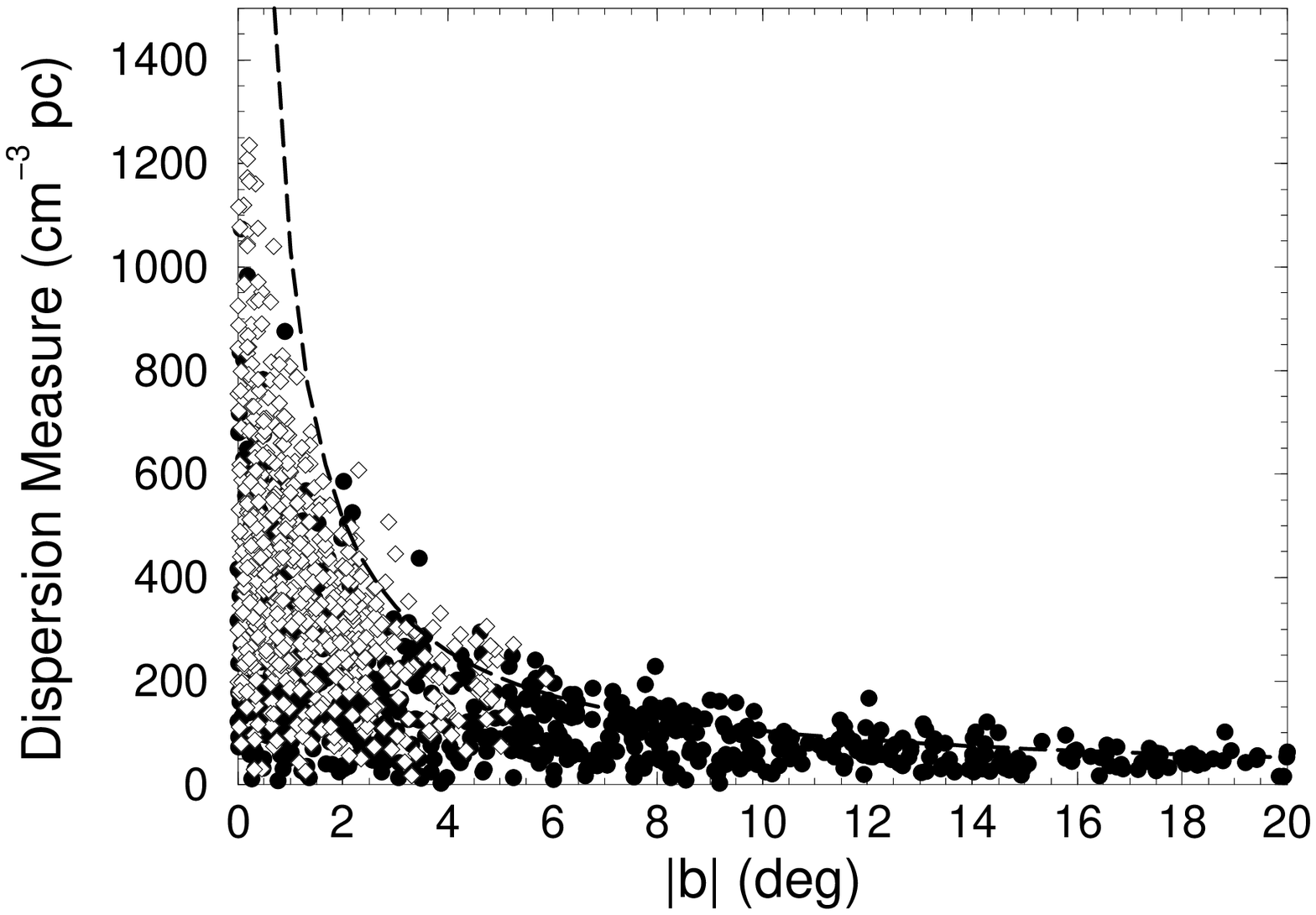,width=9cm}}

\caption{\label{f:bdm}
Observed dispersion measures as a function of the magnitude
of Galactic latitude for PMPS pulsars (open
diamonds) and all others (filled circles). The dashed
line is described by DM$=18/\sin|b|$ cm$^{-3}$ pc.}
\end{figure}

\begin{figure}

\centerline{\psfig{file=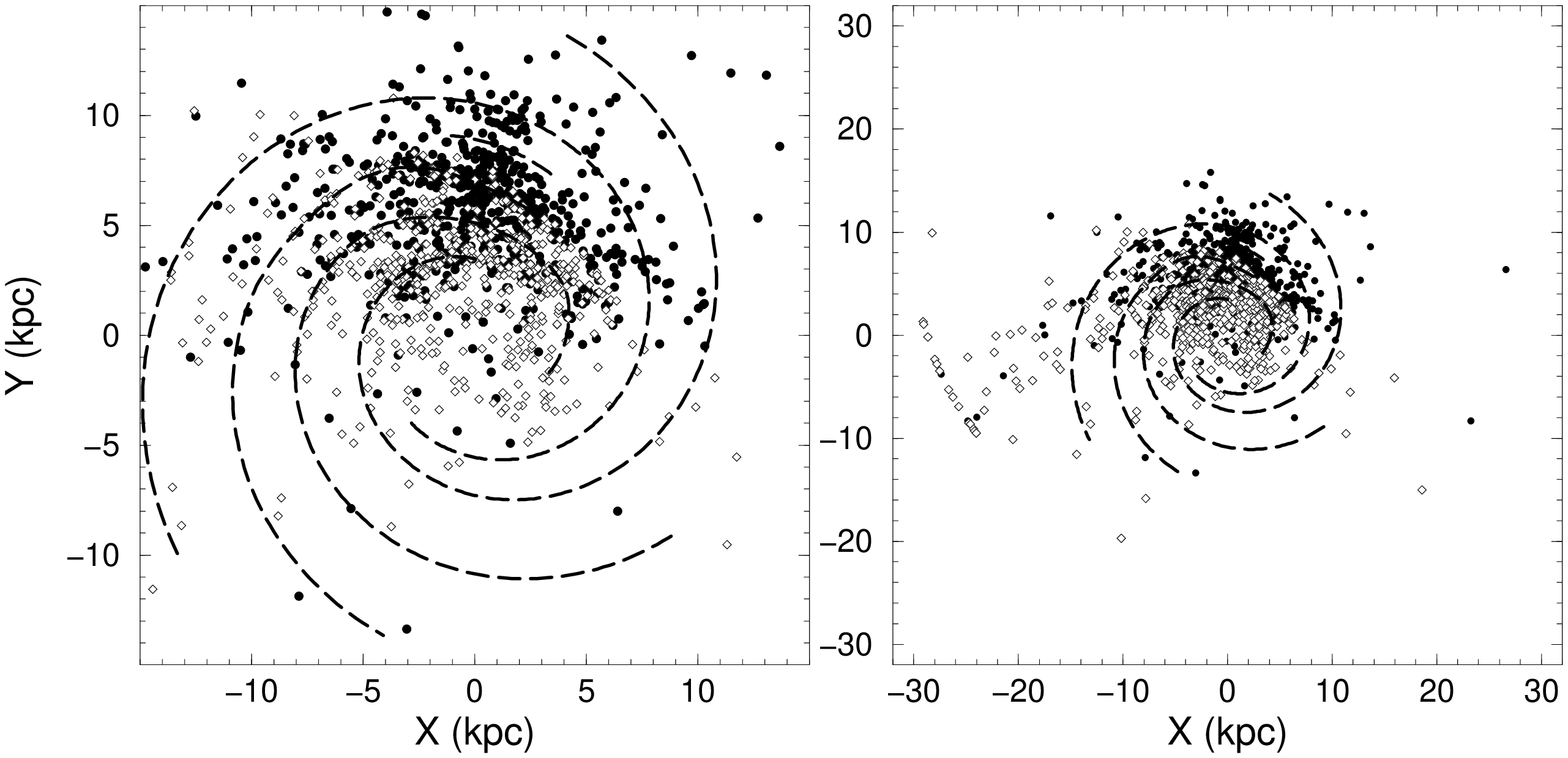,width=9cm,angle=-90}}

\caption{\label{f:planetc93} 
Location of known pulsars in the Galactic plane, $|b|\le20^\circ$,
 based on distance
estimates derived from the TC93 model. 
%Pulsars discovered in PMPS are marked by as open diamonds. 
The spiral arm
structure as used in the electron density model is indicated.
The left panel shows the inner 15 kpc around the Galactic centre,
while the right panel zooms out to demonstrate the existence of
an artificial 30-kpc ring around the Sun (located
at (0,8.5)) caused by electron deficits in the
TC93 model.}
\end{figure}

\begin{figure}

\centerline{\psfig{file=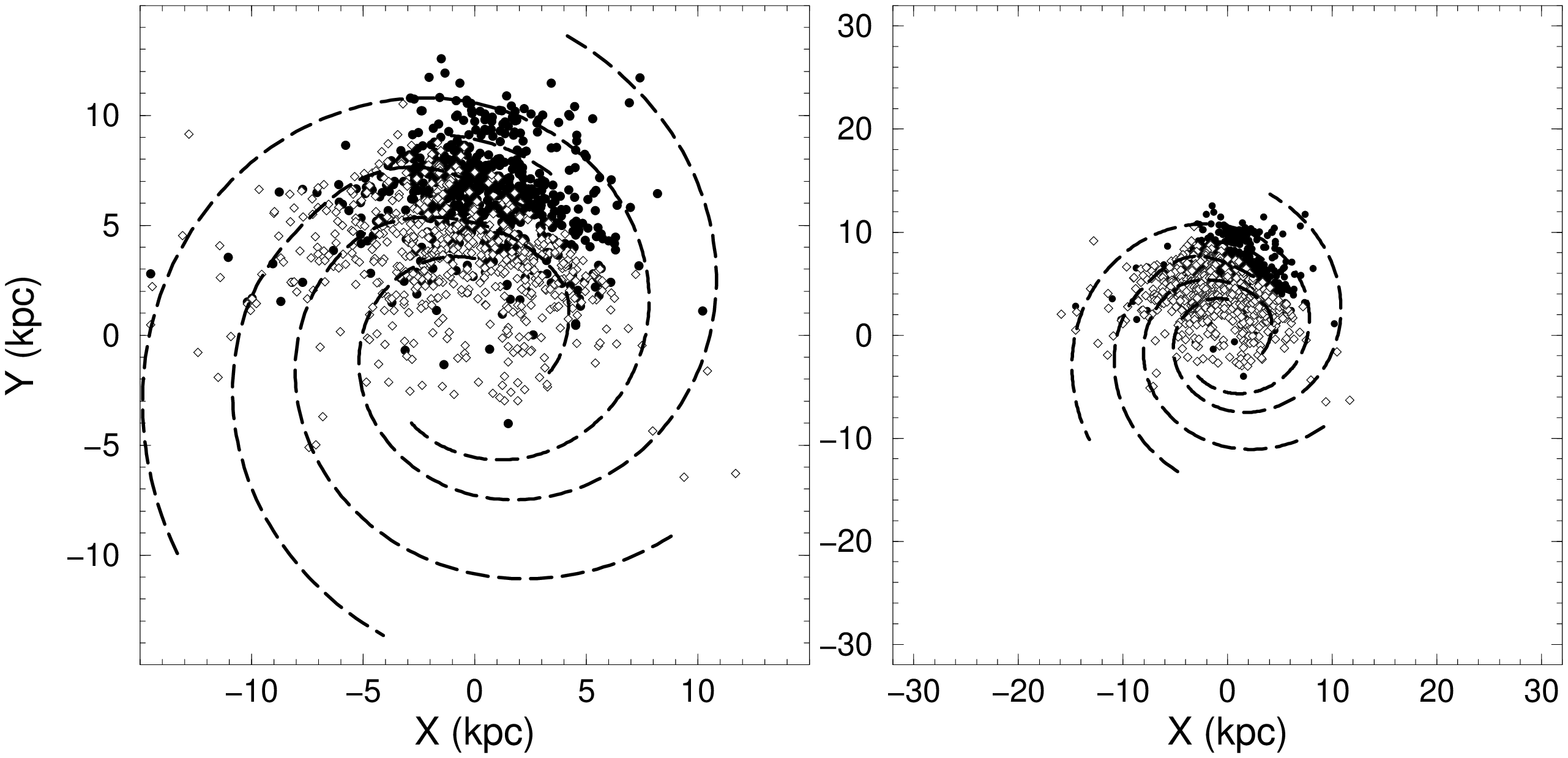,width=9cm,angle=-90}}

\caption{\label{f:planene2001} 
Location of known pulsars in the Galactic plane, $|b|\le20^\circ$,
 based on distance
estimates derived from the NE2001 model. 
%Pulsars discovered in PMPS are marked with open diamonds. 
The right 
panel demonstrates that the artificial 30-kpc ring around the Sun 
seen in the TC93 model has disappeared.}
\end{figure}

\begin{figure}

\centerline{\psfig{file=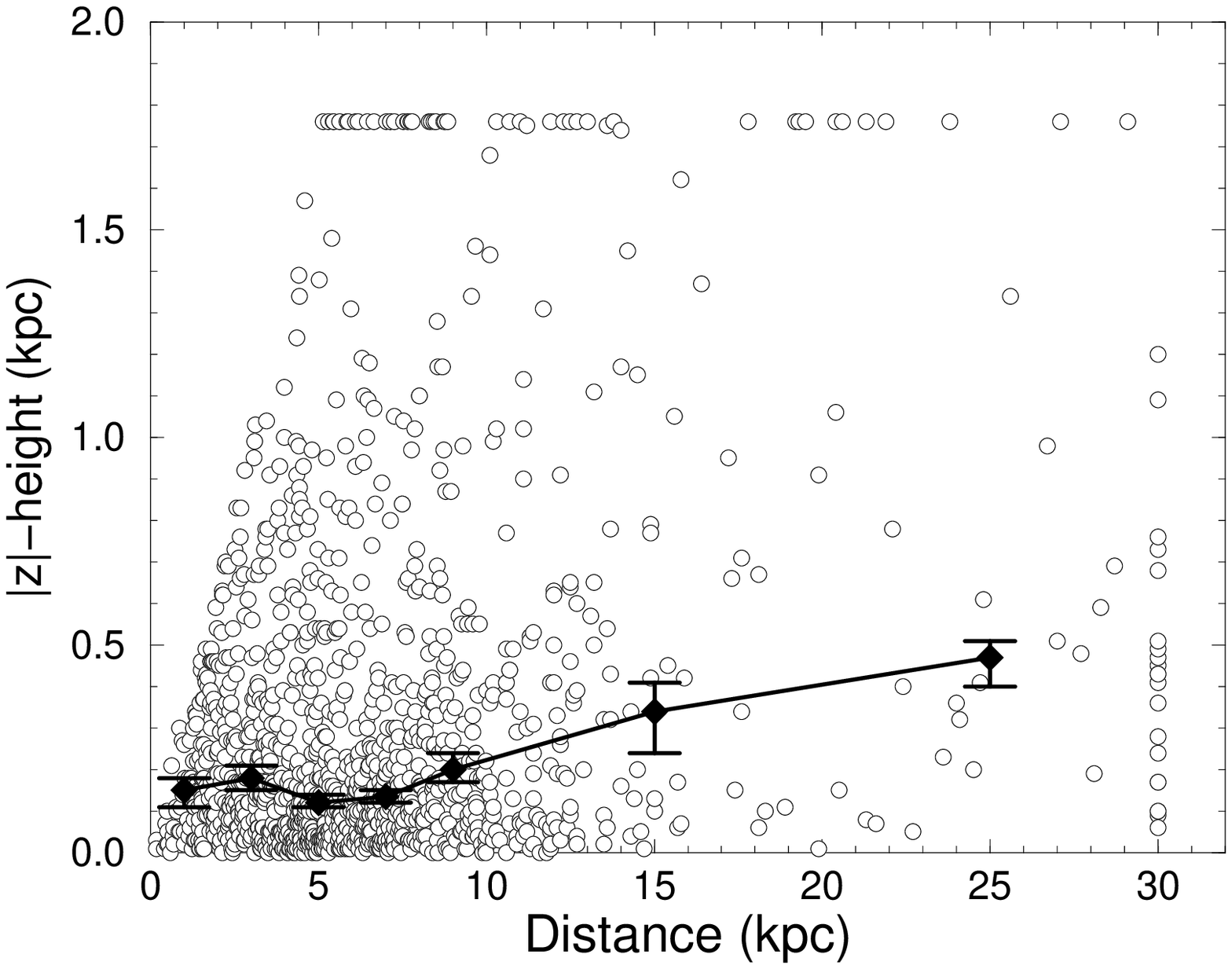,width=9cm}}

\caption{\label{f:zlold} 
Location of known pulsars with $|b|\le20^\circ$
above and below the Galactic plane 
based on distance estimates derived from the TC93 model. The pulsars
lying along a line of constant $z$-height and distance
indicate an artifact in the model. Filled diamonds represent medians
computed in 2-kpc intervals for distances below 10 kpc, and in
10-kpc intervals for larger distances, respectively.
The shown error bars are obtained from the following reasoning: The
median divides a sample of $n$ pulsars, located in a given 2-kpc
interval and sorted according to their $|z|$-height, into sub-sets of
$n/2$ pulsars with $|z|$-values larger and smaller than the median,
respectively, i.e.~$|z|_{\rm median} = |z|_{n/2}$.  We therefore
estimate an uncertainty of $\pm\sqrt{n/2}$ for the number of pulsars
in each $n/2$-sub-set. The error bars are then determined as the
differences in $|z|$-height of the median and the $\sqrt{n/2}$-th
element (counted from the median) in each sub-set. 
}
\end{figure}

\begin{figure}

\centerline{\psfig{file=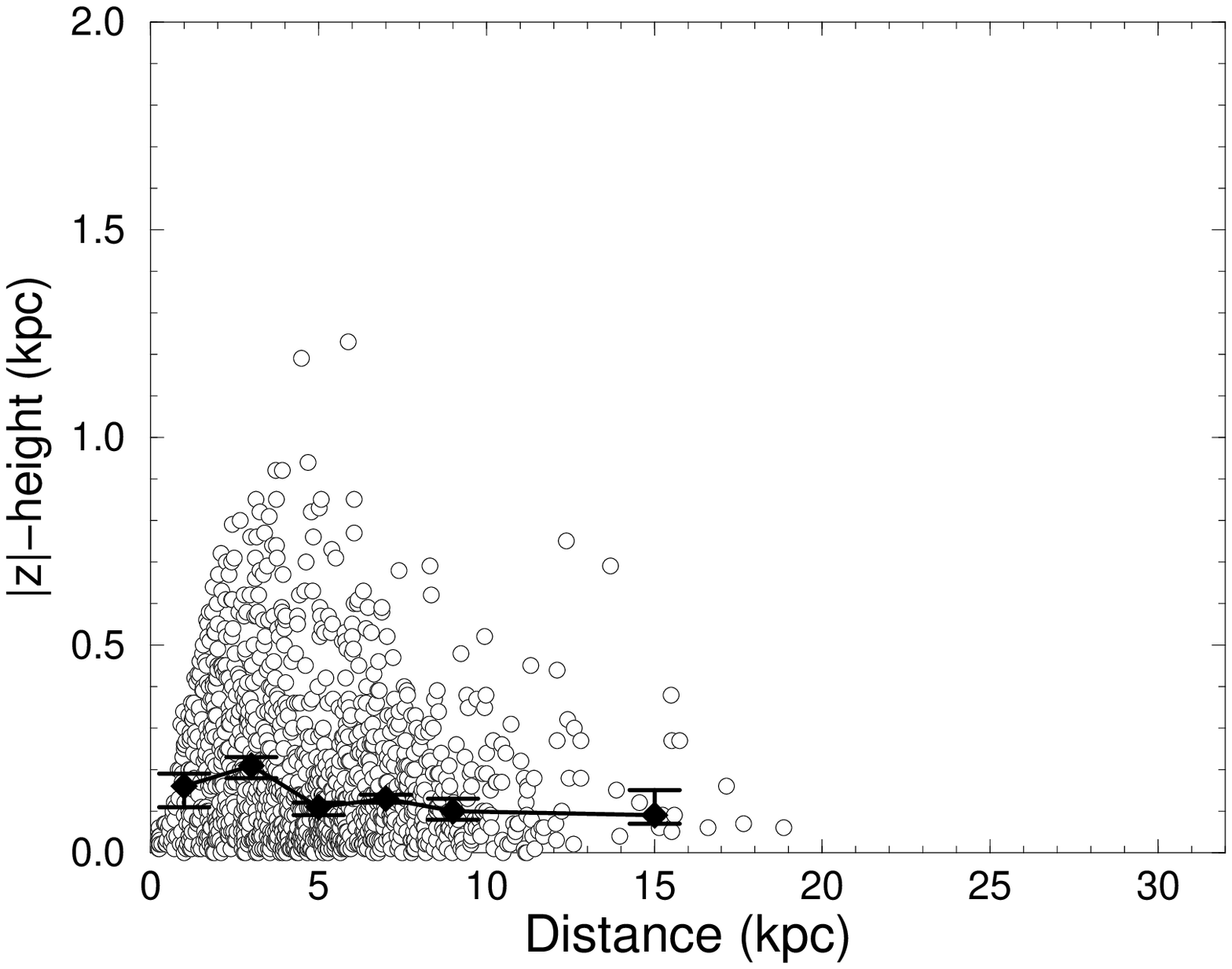,width=9cm}}

\caption{\label{f:zlnew} As previous figure, but with 
locations of known pulsars 
based on distance estimates derived from the NE2001 model. }
\end{figure}

\end{document}